\documentclass[
aps,%
12pt,%
final,%
notitlepage,%
oneside,%
onecolumn,%
nobibnotes,%
nofootinbib,%
superscriptaddress,%
noshowpacs,%
centertags]%
{revtex4}

\newcommand{\degree}{^\circ} 
\newcommand{\eq}[1]{(\ref{eq:#1})}
\newcommand{\fig}[1]{Fig.\,\ref{fig:#1}}
\newcommand{\tab}[1]{Tab.\,\ref{tab:#1}}
\DeclareMathOperator{\const}{const}
\begin{document}
\selectlanguage{english}
\title{Four years of wide-field search for nanosecond optical transients with the TAIGA-HiSCORE Cherenkov array}

\author{\firstname{A.~D.}~\surname{Panov}}
\email{panov@dec1.sinp.msu.ru}
\affiliation{%
Lomonosov Moscow State University Skobeltsyn Institute of Nuclear Physics (MSU SINP),
Leninskie gory 1(2), GSP-1, Moscow, 119991, Russia.
}
\author{\firstname{I.~I.}~\surname{Astapov}}
\affiliation{%
National Research Nuclear University MEPhI (Moscow Engineering Physics Institute),
Kashirskoe highway 31, Moscow, 115409, Russia.
}
\author{\firstname{G.~M.}~\surname{Beskin}}
\affiliation{%
Special Astrophysical Observatory, Nizhnij Arkhyz, Zelenchukskiy region, Karachai-Cherkessian Republic, 369167, Russia.
}
\author{\firstname{P.~A.}~\surname{Bezyazykov}}
\affiliation{%
Institute of Applied Physics, Irkutsk State University (API ISU), Gagarin Blvd. 20, Irkutsk,
664003, Russia.
}
\author{\firstname{A.~V.}~\surname{Blinov}}
\affiliation{%
Joint Institute for Nuclear Research, Joliot-Curie 6, Dubna, Moscow Region, 141980, Russia.
}
\author{\firstname{E.~A.}~\surname{Bonvech}}
\affiliation{%
Lomonosov Moscow State University Skobeltsyn Institute of Nuclear Physics (MSU SINP),
Leninskie gory 1(2), GSP-1, Moscow, 119991, Russia.
}
\author{\firstname{A.~N.}~\surname{Borodin}}
\affiliation{%
Joint Institute for Nuclear Research, Joliot-Curie 6, Dubna, Moscow Region, 141980, Russia.
}
\author{\firstname{N.~M.}~\surname{Budnev}}
\affiliation{%
Institute of Applied Physics, Irkutsk State University (API ISU), Gagarin Blvd. 20, Irkutsk,
664003, Russia.
}
\author{\firstname{A.~V.}~\surname{Bulan}}
\affiliation{%
Lomonosov Moscow State University Skobeltsyn Institute of Nuclear Physics (MSU SINP),
Leninskie gory 1(2), GSP-1, Moscow, 119991, Russia.
}
\author{\firstname{P.}~\surname{Busygina}}
\affiliation{%
Institute of Applied Physics, Irkutsk State University (API ISU), Gagarin Blvd. 20, Irkutsk,
664003, Russia.
}
\author{\firstname{D.~V.}~\surname{Chernov}}
\affiliation{%
Lomonosov Moscow State University Skobeltsyn Institute of Nuclear Physics (MSU SINP),
Leninskie gory 1(2), GSP-1, Moscow, 119991, Russia.
}
\author{\firstname{A.}~\surname{Chiavassa}}
\affiliation{%
Physics Department of the University of Torino and the National Institute of Nuclear
Physics INFN, 10125 Torino, Italy.
}%
\author{\firstname{A.~N.}~\surname{Dyachok}}
\affiliation{%
Institute of Applied Physics, Irkutsk State University (API ISU), Gagarin Blvd. 20, Irkutsk,
664003, Russia.
}
\author{\firstname{A.~R.}~\surname{Gafarov}}
\affiliation{%
Institute of Applied Physics, Irkutsk State University (API ISU), Gagarin Blvd. 20, Irkutsk,
664003, Russia.
}
\author{\firstname{A.~Yu.}~\surname{Garmash}}
\affiliation{%
Novosibirsk State University, Pirogova 1, Novosibirsk, 630090, Russia.
}
\affiliation{%
Budker Institute of Nuclear Physics of the Siberian Branch of the Russian Academy of
Sciences, Lavrentyev Prosp. 11, Novosibirsk, 630090, Russia.
}
\author{\firstname{V.~M.}~\surname{Grebenyuk}}
\affiliation{%
Joint Institute for Nuclear Research, Joliot-Curie 6, Dubna, Moscow Region, 141980, Russia.
}
\author{\firstname{E.~O.}~\surname{Gress}}
\affiliation{%
Institute of Applied Physics, Irkutsk State University (API ISU), Gagarin Blvd. 20, Irkutsk,
664003, Russia.
}
\author{\firstname{O.~A.}~\surname{Gress}}
\affiliation{%
Institute of Applied Physics, Irkutsk State University (API ISU), Gagarin Blvd. 20, Irkutsk,
664003, Russia.
}
\author{\firstname{T.~I.}~\surname{Gress}}
\affiliation{%
Institute of Applied Physics, Irkutsk State University (API ISU), Gagarin Blvd. 20, Irkutsk,
664003, Russia.
}
\author{\firstname{A.~A.}~\surname{Grinyuk}}
\affiliation{%
Joint Institute for Nuclear Research, Joliot-Curie 6, Dubna, Moscow Region, 141980, Russia.
}
\author{\firstname{O.~G.}~\surname{Grishin}}
\affiliation{%
Institute of Applied Physics, Irkutsk State University (API ISU), Gagarin Blvd. 20, Irkutsk,
664003, Russia.
}
\author{\firstname{A.~L.}~\surname{Ivanova}}
\affiliation{%
Institute of Applied Physics, Irkutsk State University (API ISU), Gagarin Blvd. 20, Irkutsk,
664003, Russia.
}%
\author{\firstname{A.~D.}~\surname{Ivanova}}
\affiliation{%
Institute of Applied Physics, Irkutsk State University (API ISU), Gagarin Blvd. 20, Irkutsk,
664003, Russia.
}%
\author{\firstname{M.~A.}~\surname{Ilyushin}}
\affiliation{%
Institute of Applied Physics, Irkutsk State University (API ISU), Gagarin Blvd. 20, Irkutsk,
664003, Russia.
}%
\author{\firstname{N.~N.}~\surname{Kalmykov}}
\affiliation{%
Lomonosov Moscow State University Skobeltsyn Institute of Nuclear Physics (MSU SINP),
Leninskie gory 1(2), GSP-1, Moscow, 119991, Russia.
}
\author{\firstname{V.~V.}~\surname{Kindin}}
\affiliation{%
Institute of Applied Physics, Irkutsk State University (API ISU), Gagarin Blvd. 20, Irkutsk,
664003, Russia.
}
\author{\firstname{S.~N.}~\surname{Kiryuhin}}
\affiliation{%
Institute of Applied Physics, Irkutsk State University (API ISU), Gagarin Blvd. 20, Irkutsk,
664003, Russia.
}
\author{\firstname{\relax}~\surname{\fbox{R.~P.~Kokoulin}}}
\affiliation{%
National Research Nuclear University MEPhI (Moscow Engineering Physics Institute),
Kashirskoe highway 31, Moscow, 115409, Russia.
}
\author{\firstname{K.~G.}~\surname{Kompaniets}}
\affiliation{%
Institute of Applied Physics, Irkutsk State University (API ISU), Gagarin Blvd. 20, Irkutsk,
664003, Russia.
}
\author{\firstname{E.~E.}~\surname{Korosteleva}}
\affiliation{%
Lomonosov Moscow State University Skobeltsyn Institute of Nuclear Physics (MSU SINP),
Leninskie gory 1(2), GSP-1, Moscow, 119991, Russia.
}
\author{\firstname{V.~A.}~\surname{Kozhin}}
\affiliation{%
Lomonosov Moscow State University Skobeltsyn Institute of Nuclear Physics (MSU SINP),
Leninskie gory 1(2), GSP-1, Moscow, 119991, Russia.
}
\author{\firstname{E.~A.}~\surname{Kravchenko}}
\affiliation{%
Novosibirsk State University, Pirogova 1, Novosibirsk, 630090, Russia.
}
\affiliation{%
Budker Institute of Nuclear Physics of the Siberian Branch of the Russian Academy of
Sciences, Lavrentyev Prosp. 11, Novosibirsk, 630090, Russia.
}
\author{\firstname{L.~A.}~\surname{Kuzmichev}}
\email{kuz@dec1.sinp.msu.ru}
\affiliation{%
Lomonosov Moscow State University Skobeltsyn Institute of Nuclear Physics (MSU SINP),
Leninskie gory 1(2), GSP-1, Moscow, 119991, Russia.
}
\author{\firstname{A.~P.}~\surname{Kryukov}}
\affiliation{%
Lomonosov Moscow State University Skobeltsyn Institute of Nuclear Physics (MSU SINP),
Leninskie gory 1(2), GSP-1, Moscow, 119991, Russia.
}
\author{\firstname{A.~A.}~\surname{Lagutin}}
\affiliation{%
Altai State University, Lenina 61, Barnaul, 656049, Russia.
}
\author{\firstname{M.~V.}~\surname{Lavrova}}
\affiliation{%
Joint Institute for Nuclear Research, Joliot-Curie 6, Dubna, Moscow Region, 141980, Russia.
}
\author{\firstname{Yu.}~\surname{Lemeshev}}
\affiliation{%
Institute of Applied Physics, Irkutsk State University (API ISU), Gagarin Blvd. 20, Irkutsk,
664003, Russia.
}
\author{\firstname{B.~K.}~\surname{Lubsandorzhiev}}
\affiliation{%
Institute for Nuclear Research of the Russian Academy of Sciences, 60th October
Anniversary 7a, 117312, Moscow, Russia.
}
\author{\firstname{N.~B.}~\surname{Lubsandorzhiev}}
\affiliation{%
Lomonosov Moscow State University Skobeltsyn Institute of Nuclear Physics (MSU SINP),
Leninskie gory 1(2), GSP-1, Moscow, 119991, Russia.
}
\author{\firstname{A.~D.}~\surname{Lukanov}}
\affiliation{%
Institute for Nuclear Research of the Russian Academy of Sciences, 60th October
Anniversary 7a, 117312, Moscow, Russia.
}
\author{\firstname{S.~D.}~\surname{Malakhov}}
\affiliation{%
Institute of Applied Physics, Irkutsk State University (API ISU), Gagarin Blvd. 20, Irkutsk,
664003, Russia.
}
\author{\firstname{R.~R.}~\surname{Mirgazov}}
\affiliation{%
Institute of Applied Physics, Irkutsk State University (API ISU), Gagarin Blvd. 20, Irkutsk,
664003, Russia.
}
\author{\firstname{R.~D.}~\surname{Monkhoev}}
\affiliation{%
Institute of Applied Physics, Irkutsk State University (API ISU), Gagarin Blvd. 20, Irkutsk,
664003, Russia.
}
\author{\firstname{E.~A.}~\surname{Okyneva}}
\affiliation{%
Lomonosov Moscow State University Skobeltsyn Institute of Nuclear Physics (MSU SINP),
Leninskie gory 1(2), GSP-1, Moscow, 119991, Russia.
}
\author{\firstname{E.~A.}~\surname{Osipova}}
\affiliation{%
Lomonosov Moscow State University Skobeltsyn Institute of Nuclear Physics (MSU SINP),
Leninskie gory 1(2), GSP-1, Moscow, 119991, Russia.
}
\author{\firstname{A.~L.}~\surname{Pakhorukov}}
\affiliation{%
Institute of Applied Physics, Irkutsk State University (API ISU), Gagarin Blvd. 20, Irkutsk,
664003, Russia.
}
\author{\firstname{A.}~\surname{Pan}}
\affiliation{%
Joint Institute for Nuclear Research, Joliot-Curie 6, Dubna, Moscow Region, 141980, Russia.
}%
\author{\firstname{L.~V.}~\surname{Pankov}}
\affiliation{%
Institute of Applied Physics, Irkutsk State University (API ISU), Gagarin Blvd. 20, Irkutsk,
664003, Russia.
}
\author{\firstname{A.~A.}~\surname{Petrukhin}}
\affiliation{%
National Research Nuclear University MEPhI (Moscow Engineering Physics Institute),
Kashirskoe highway 31, Moscow, 115409, Russia.
}
\author{\firstname{D.~A.}~\surname{Podgrudkov}}
\affiliation{%
Lomonosov Moscow State University Skobeltsyn Institute of Nuclear Physics (MSU SINP),
Leninskie gory 1(2), GSP-1, Moscow, 119991, Russia.
}
\author{\firstname{I.~A.}~\surname{Poddubny}}
\affiliation{%
Institute of Applied Physics, Irkutsk State University (API ISU), Gagarin Blvd. 20, Irkutsk,
664003, Russia.
}
\author{\firstname{E.~G.}~\surname{Popova}}
\affiliation{%
Lomonosov Moscow State University Skobeltsyn Institute of Nuclear Physics (MSU SINP),
Leninskie gory 1(2), GSP-1, Moscow, 119991, Russia.
}
\author{\firstname{E.~B.}~\surname{Postnikov}}
\affiliation{%
Lomonosov Moscow State University Skobeltsyn Institute of Nuclear Physics (MSU SINP),
Leninskie gory 1(2), GSP-1, Moscow, 119991, Russia.
}
\author{\firstname{V.~V.}~\surname{Prosin}}
\affiliation{%
Lomonosov Moscow State University Skobeltsyn Institute of Nuclear Physics (MSU SINP),
Leninskie gory 1(2), GSP-1, Moscow, 119991, Russia.
}%
\author{\firstname{A.~A.}~\surname{Pushnin}}
\affiliation{%
Institute of Applied Physics, Irkutsk State University (API ISU), Gagarin Blvd. 20, Irkutsk,
664003, Russia.
}%
\author{\firstname{R.~I.}~\surname{Raikin}}
\affiliation{%
Altai State University, Lenina 61, Barnaul, 656049, Russia.
}%
\author{\firstname{A.~Yu.}~\surname{Razumov}}
\affiliation{%
Lomonosov Moscow State University Skobeltsyn Institute of Nuclear Physics (MSU SINP),
Leninskie gory 1(2), GSP-1, Moscow, 119991, Russia.
}%
\author{\firstname{E.}~\surname{Rjabov}}
\affiliation{%
Institute of Applied Physics, Irkutsk State University (API ISU), Gagarin Blvd. 20, Irkutsk,
664003, Russia.
}
\author{\firstname{G.~I.}~\surname{Rubtsov}}
\affiliation{%
Institute for Nuclear Research of the Russian Academy of Sciences, 60th October
Anniversary 7a, 117312, Moscow, Russia.
}%
\author{\firstname{V.~S.}~\surname{Samoliga}}
\affiliation{%
Institute of Applied Physics, Irkutsk State University (API ISU), Gagarin Blvd. 20, Irkutsk,
664003, Russia.
}
\author{\firstname{A.~V.}~\surname{Shaikovsky}}
\affiliation{%
Joint Institute for Nuclear Research, Joliot-Curie 6, Dubna, Moscow Region, 141980, Russia.
}
\author{\firstname{A.~Yu.}~\surname{Sidorenkov}}
\affiliation{%
Institute for Nuclear Research of the Russian Academy of Sciences, 60th October
Anniversary 7a, 117312, Moscow, Russia.
}
\author{\firstname{A.~A.}~\surname{Silaev}}
\affiliation{%
Lomonosov Moscow State University Skobeltsyn Institute of Nuclear Physics (MSU SINP),
Leninskie gory 1(2), GSP-1, Moscow, 119991, Russia.
}
\author{\firstname{A. A.}~\surname{Silaev (junior)}}
\affiliation{%
Lomonosov Moscow State University Skobeltsyn Institute of Nuclear Physics (MSU SINP),
Leninskie gory 1(2), GSP-1, Moscow, 119991, Russia.
}
\author{\firstname{A.~V.}~\surname{Skurikhin}}
\affiliation{%
Lomonosov Moscow State University Skobeltsyn Institute of Nuclear Physics (MSU SINP),
Leninskie gory 1(2), GSP-1, Moscow, 119991, Russia.
}
\author{\firstname{I.}~\surname{Satyshev}}
\affiliation{%
Joint Institute for Nuclear Research, Joliot-Curie 6, Dubna, Moscow Region, 141980, Russia.
}
\author{\firstname{A.~V.}~\surname{Sokolov}}
\affiliation{%
Novosibirsk State University, Pirogova 1, Novosibirsk, 630090, Russia.
}
\affiliation{%
Budker Institute of Nuclear Physics of the Siberian Branch of the Russian Academy of
Sciences, Lavrentyev Prosp. 11, Novosibirsk, 630090, Russia.
}
\author{\firstname{L.~G.}~\surname{Sveshnikova}}
\affiliation{%
Lomonosov Moscow State University Skobeltsyn Institute of Nuclear Physics (MSU SINP),
Leninskie gory 1(2), GSP-1, Moscow, 119991, Russia.
}
\author{\firstname{V.~A.}~\surname{Tabolenko}}
\affiliation{%
Institute of Applied Physics, Irkutsk State University (API ISU), Gagarin Blvd. 20, Irkutsk,
664003, Russia.
}
\author{\firstname{A.~B.}~\surname{Tanaev}}
\affiliation{%
Institute of Applied Physics, Irkutsk State University (API ISU), Gagarin Blvd. 20, Irkutsk,
664003, Russia.
}
\author{\firstname{M.}~\surname{Ternovoy}}
\affiliation{%
Institute of Applied Physics, Irkutsk State University (API ISU), Gagarin Blvd. 20, Irkutsk,
664003, Russia.
}
\author{\firstname{\relax}~\surname{\fbox{L.~G.~Tkachev}}}
\affiliation{%
Joint Institute for Nuclear Research, Joliot-Curie 6, Dubna, Moscow Region, 141980, Russia.
}
\affiliation{%
Dubna State University, Universitetskaya 19, Dubna, Moscow region, 141982, Russia.
}
\author{\firstname{N.}~\surname{Ushakov}}
\affiliation{%
Institute for Nuclear Research of the Russian Academy of Sciences, 60th October
Anniversary 7a, 117312, Moscow, Russia.
}
\author{\firstname{P.~A.}~\surname{Volchugov}}
\affiliation{%
Lomonosov Moscow State University Skobeltsyn Institute of Nuclear Physics (MSU SINP),
Leninskie gory 1(2), GSP-1, Moscow, 119991, Russia.
}
\author{\firstname{N.~V.}~\surname{Volkov}}
\affiliation{%
Altai State University, Lenina 61, Barnaul, 656049, Russia.
}
\author{\firstname{D.~M.}~\surname{Voronin}}
\affiliation{%
Institute for Nuclear Research of the Russian Academy of Sciences, 60th October
Anniversary 7a, 117312, Moscow, Russia.
}%
\author{\firstname{I.~I.}~\surname{Yashin}}
\affiliation{%
National Research Nuclear University MEPhI (Moscow Engineering Physics Institute),
Kashirskoe highway 31, Moscow, 115409, Russia.
}%
\author{\firstname{A.~V.}~\surname{Zagorodnikov}}
\affiliation{%
Institute of Applied Physics, Irkutsk State University (API ISU), Gagarin Blvd. 20, Irkutsk,
664003, Russia.
}%
\author{\firstname{D.~P.}~\surname{Zhurov}}
\affiliation{%
Institute of Applied Physics, Irkutsk State University (API ISU), Gagarin Blvd. 20, Irkutsk,
664003, Russia.
}%
\author{\firstname{V.~N.}~\surname{Zirakashvili}}
\affiliation{%
Pushkov Institute of Terrestrial Magnetism, Ionosphere and Radio Wave Propagation of the
Siberian Branch of the Russian Academy of Sciences (IZMIRAN), Kaluzhskoe highway 4,
Moscow, Troitsk, 108840, Russia.
}%

\begin{abstract}
It has been previously demonstrated [Panov et al. Physics of Atomic Nuclei 84(2021)1037] that the TAIGA-HiSCORE Cherenkov array, originally built for cosmic ray physics and ultrahigh-energy gamma-ray astronomy studies using the extensive air shower method, can be used in conventional optical astronomy for wide-field searches for rare nanosecond optical transients of astrophysical origin. The FOV of the facility is on the scale of 1~ster, and it is capable of detecting very rare transients in the visible light range with fluxes greater than approximately 3000~quanta/m$^2$/10~ns (10~ns is the apparatus integration time) and pulse durations of 10\,ns. Among the potential sources of distant nanosecond optical transients are the evaporation of primary black holes, magnetic reconnection in the accretion disks of black holes, and signals from distant lasers of extraterrestrial civilizations. The paper describes the methods and results of the search for optical transients using the TAIGA-HiSCORE Cherenkov array from 2018 to 2022 (four winter seasons of data collection). No reliable astrophysical candidates for optical transients were found. We set an upper bound on the flux of the searched events as $\sim 1\times10^{-3}$\,events/ster/h.
\end{abstract}

\maketitle

\section{Introduction}


The TAIGA (Tunka Advanced Instrument for cosmic ray physics and Gamma-ray Astronomy) astrophysical complex is designed to study cosmic ray physics and gamma-ray astronomy at high and ultra-high energies using a technique based on the observation of extensive air showers (EAS) \cite{TAIGA-NIM-2017A,TAIGA-NIM-2017B,TAIGA-NIM-2020A,TAIGA-NIM-2020B}. The complex is located in the Tunka Valley, 50 km from the Lake Baikal. Most of the technologies implemented in the TAIGA complex are related to the detection of the Cherenkov radiation of the EAS. The detection of Cherenkov light of EAS requires clear and dark moonless nights, which in the area where the complex is located are mainly from late autumn to early spring, so the work of the complex has a seasonal character. Below we will speak about <<winter>> seasons of operation of the TAIGA complex, which actually includes the period usually from October to April.


The complex includes several instruments, including the TAIGA-HiSCORE wide-aperture Cherenkov array. The HiSCORE (High Sensitivity COsmic Rays and Gamma Explorer) array consists of Cherenkov stations, each of which contains four PMTs with a total entrance pupil area of about 0.5~m${}^2$ and a field of view (FOV) on the order of a steradian. The exact size of the FOV is difficult to determine because the FOV has no sharp boundaries, the position of which also depends on the brightness of the flash. For quantitative estimations, we will assume an FOV value of 0.6~ster, as was done in \cite{PANOV-2021}.  All stations are oriented in the same way, but their orientation may change from season to season. The stations form a network with a spacing of 106~m and is currently represented by 120 stations spread over an area of more than 1~km${}^2$. The design and construction details of the HiSCORE array are presented in the articles \cite{HISCORE-2011,TAIGA-NIM-2017B}. The working principle of the TAIGA-HiSCORE array is based on the idea presented in \cite{HISCORE-2011}. The optical stations measure the amplitudes of the light signal caused by the Cherenkov radiation of an EAS, as well as the time evolution of the Cherenkov light front reconstructed from the sequence of station triggering. It is possible from these data to reconstruct the arrival direction of a cosmic particle (charged particle or gamma ray) with precision about 0.1$\degree$, the position of the shower axis, its initial energy and some other important characteristics of the EAS.

The characteristic duration of EAS Cherenkov light pulses is a few nanoseconds to tens of nanoseconds, so the HiSCORE detector stations are designed to detect such flashes. The HiSCORE integration time is about 10~ns. The stations have a detection threshold of about 3000 quanta of green-band light per square meter per integration time. In principle, the arraye is capable of detecting not only pulses of EAS Cherenkov light, but any nanosecond flashes within the FOV of HiSCORE and above the threshold. In the paper \cite{PANOV-2021} it was practically demonstrated that the flash of a distant point source (hereafter \emph{optical transient}) can be distinguished from the passage of the EAS Cherenkov front by a number of characteristics of the light front. Importantly, optical transient events should easily pass the primary event triggers in the HiSCORE core working methodology, so that all potential candidates for transient events enter a common database along with EAS events. To search for transients in the HiSCORE data, it is only necessary to apply special algorithms to the processing of the events in the common database, so that the search for optical transients can be implemented in a companion mode, without requiring any additional hardware settings and without interfering in any way with the implementation of the main scientific programs of the TAIGA astrophysical complex. Thus, the HiSCORE Cherenkov array, originally adapted to the study of cosmic ray physics and gamma-ray astronomy, can be used relatively easily to solve problems related to conventional optical astronomy.


The idea of using Cherenkov telescopes to search for optical transients of astrophysical origin was proposed already in the very early 2000s \cite{CH-BESKIN-2001,CH-SETI-EICHLER-BESKIN-2001}. After that, the idea was repeatedly discussed in different aspects and some search programs were implemented \cite{CH-SETI-HOLDER-2005,CH-GRIFFIN-2011,CH-BARTOS-2014,CH-BARTOS-2018,CH-MAGIC-2018,CH-MAGIC-2019,CH-MAGIC-2023,CH-VERITAS-2021,CH-VERITAS-2023}. However, all these papers have considered and discussed the use of Imaging Atmospheric Cherenkov Telescopes (IACT), which have a very small FOV, measured in angles of only units of degrees. Such instruments are not well suited to search for very rare optical transients that may come from a direction unknown in advance. In the present work, which is a continuation of \cite{PANOV-2021}, a completely different type of instrument is used for the first time to search for optical transients. It is very important in this context that the HiSCORE Cherenkov array, in contrast to any IACT, has a very large FOV, which makes it well suited to search for very rare events. Interestingly, the HiSCORE array is the only optical astronomical instrument to date that does not use imaging optics. Instead, the direction to the source is determined by the response times of the optical stations, each of which does not build up an optical image, but looks at a very large part of the sky at once.

In the paper \cite{PANOV-2021}, we described the methods for searching for optical transients and the results of the first observing season (winter 2018--2019) of the HiSCORE array operation. In \cite{PANOV-2021} we managed to almost completely remove the EAS background for events near the zenith (zenith angles $\theta < 40\degree$), but in the interval $40\degree < \theta < 60\degree$ the background remained quite high. Angles $\theta > 60\degree$ were outside the FOV of the HiSCORE array. Due to low background it was possible to search for single optical transient candidates in the region of angles $\theta < 40\degree$ (no candidates have been detected, except for the CALIPSO satellite signals). In the range of angles $40\degree < \theta < 60\degree$, the signature of astrophysical sources could only be the repetition of signals from the same point in the sky within the errors of the angular measurements. No such statistically reliable repeaters were found. In the present work we analyze the data from three new seasons of HiSCORE array observations (2019--2020, 2020--2021, 2021--2022). The search methods have been significantly improved. Several very strong new filters of the astrophysical optical transient have been added to those used in the processing of the 2018---2019 season data. The EAS background has been reduced to nearly zero over the entire field of view. The present paper describes the techniques used and the results of the search for nanosecond optical transients in three new HiSCORE seasons 2019--2020, 2020--2021, 2021--2022. Thus, the processing of four HiSCORE seasons of the nanosecond optical transient search has now been completed.

\section{Physical motivation of the search for optical transients of astrophysical origin in the nanosecond range}


An interesting question is what kind of physics might be behind the search for nanosecond optical transients with FOVs on the scale of one steradian. We distinguish four categories of problems.


1. First, HiSCORE searches for optical signals in the range of experimental parameters where astronomical studies have never been performed before. The novelty lies in the combination of nanosecond signal durations and a very large FOV (one steradian scale) for optical astronomy. This makes it possible to search for very rare events in the sky. When an experimental method penetrates into the previously inaccessible parameter region one can always expect to detect something completely unexpected. In fact, exactly this scenario has already worked. In the HiSCORE 2018--2019 winter season data, the lidar signals from the CALIPSO satellite were unexpectedly detected thanks to the implementation of the distant optical transient search technique with HiSCORE array \cite{PANOV-2021}. While the CALIPSO optical pulses are certainly not signals from a distant astrophysical source, they have proven to be extremely useful for calibrating all Cherenkov techniques, both on the HiSCORE array itself and with other instruments within the TAIGA complex. The CALIPSO satellite flies over the HiSCORE array at an altitude of about 700 km, so the optical front of the lidar appears very flat, similar to that of a point source located at infinity. Therefore, the CALIPSO lidar signal proved to be useful not only for calibrating the Cherenkov techniques, but also for calibrating and testing the distant optical transient search technique itself.


2. If we are talking about real astrophysical optical transients, then for an incoherent light source to produce a flash in the range of durations from units to a few tens of nanoseconds, it should have a size of no more than a few tens of meters. Such a source may be the evaporation of the remnant of a primordial black hole (PBH).


Currently, we can observe a significant increase of interest in the physics of PBHs \cite{KHLOPOV-2010,DOLGOV-POSTNOV-2016,DOLGOV-2024,CARR-2024A,CARR-2024B}. The ideas about the possibility of the existence of PBHs, their origin, abundance, and mass spectrum are still quite controversial. However, the possibility that PBHs solve three astrophysical mysteries at once is already being discussed: first, the existence of giant black holes at the centers of all large galaxies, second, the unexpectedly early formation of galaxies according to JWST and HST \cite{DOLGOV-2024}, third, the unusual parameters of merging black holes, which according to gravitational telescope data bear little resemblance to black holes of stellar origin \cite{DOLGOV-POSTNOV-2016}. In addition, the possibility that a substantial or even large fraction of dark matter is represented by PBH is discussed \cite{CARR-2020,CARR-2021}. The physics of PBH has been linked to fundamental problems beyond the standard models of particle physics and cosmology: the possible connection of PBH formation to extra dimensions \cite{RUBIN-2022}, and many others, are discussed. This stimulates interest in the physics of PBH.  Many hundreds of papers dealing with PBH physics have been published in the last decade and a half, and we are not in a position to give an overview of them in this paper.


The initial mass distribution function of PBHs depends very much on the model of their formation\cite{CARR-2021}, and at present almost any assumptions about the allowed initial masses of PBHs are acceptable. Every black hole evaporates due to Hawking radiation, and if the initial mass of the PBH was not too large, it could have evaporated during the 13.8~billion~year lifetime of the Universe. In addition, some PBHs may be evaporating in modern times. The physics of evaporating black holes is of great interest because the final phases of evaporation must be described by the currently unknown theory of quantum gravity. In fact, observing the evaporation of PBHs may be the only experimental way to observe strong effects of quantum gravity. Evaporating black holes may also be relevant for solving the well-known cosmological puzzle related to the threefold excess of the isotope $^7$Li in primary nucleosynthesis \cite{CARR-2021}.

As shown by MacGibbon \cite{MACGIBBON-1991}, the lifetime of a Schwarzschild black hole with respect to the evaporation process as a function of its initial mass $M$ can be written as
\begin{equation}
 \tau(M) = 1.19\times10^3 \frac{G^2M^3}{\hbar c^4 f_L(M)},
 \label{eq:TauM}
\end{equation}
whence it follows
\begin{equation}
M(\tau) = \sqrt[3]{\tau\,\frac{\hbar c^4 f_L(M)}{1.19\times10^3}G^2}.
 \label{eq:MTau}
\end{equation}
Here, all quantities are expressed in the unit system (g, cm, c), and $f_L(M)$ is a statistical factor that depends on the number of species of particles emitted. The instantaneous value of the number of species of emitted particles grows with increasing Hawking temperature and decreasing black hole mass, so the corresponding statistical factor should be considered as a function of temperature $f(T)$. In the equation \eq{TauM}, the factor $f_L(M)$ has the meaning of the statistical factor $f(T)$ averaged over the whole lifetime of a black hole with initial mass $M$. According to \cite{CARR-2014}, for black holes evaporating in the present epoch, $f_L \approx 1.9$, from which it is easy to find from the formula \eq{MTau} that the initial mass of black holes evaporating now is $M_* \approx 5.1\times10^{14}$\,g.

Since the Hawking temperature is inversely proportional to the mass of the black hole, the evaporation process accelerates with mass loss and, for a PBH with mass $M_*$, may end in a powerful short explosion in the present-day universe. To obtain the expected time profile of such an explosion, one can again try to use the formula \eq{MTau}. However, there is now a fundamental ambiguity in the definition of the factor $f_L(M)$. For example, if one tries to describe the last microsecond of a black hole's evaporation, the mass of the black hole at the very beginning of that microsecond will be on the order of 25 tons, according to very rough estimates. The Hawking temperature is given by the formula
\begin{equation}
 kT = \frac{\hbar c^3}{8\pi GM},
 \label{eq:kT}
\end{equation}
and as can be seen from this formula, 25 tons corresponds to a temperature on the order of $5\times10^4$\,TeV. What factor $f(T)$ corresponds to such a temperature is unknown, simply due to our ignorance of physics at such high energies. As the mass of the black hole remnant decreases, the temperature will become even higher, and the radiation process will eventually include essentially all fundamental particles with masses up to Planckian. It is not known how many species of such particles there are, so it is impossible to make reliable estimates of the factor $f_L(M)$.  The statistical factor $f_{SM} = 15.35$ corresponds to all particles of the Standard Model \cite{CARR-2014}, so for a qualitative assessment of the picture of the last microsecond of black hole evaporation it seems reasonable to take $f_L \sim 100$ for all times between microsecond and zero. Note that in the formula \eq{MTau} the factor $f_L$ is included under the square root, which reduces the influence of the uncertainty $f_L$ on the final estimates.

\begin{figure}
 \begin{center}
 \includegraphics[width=\textwidth]{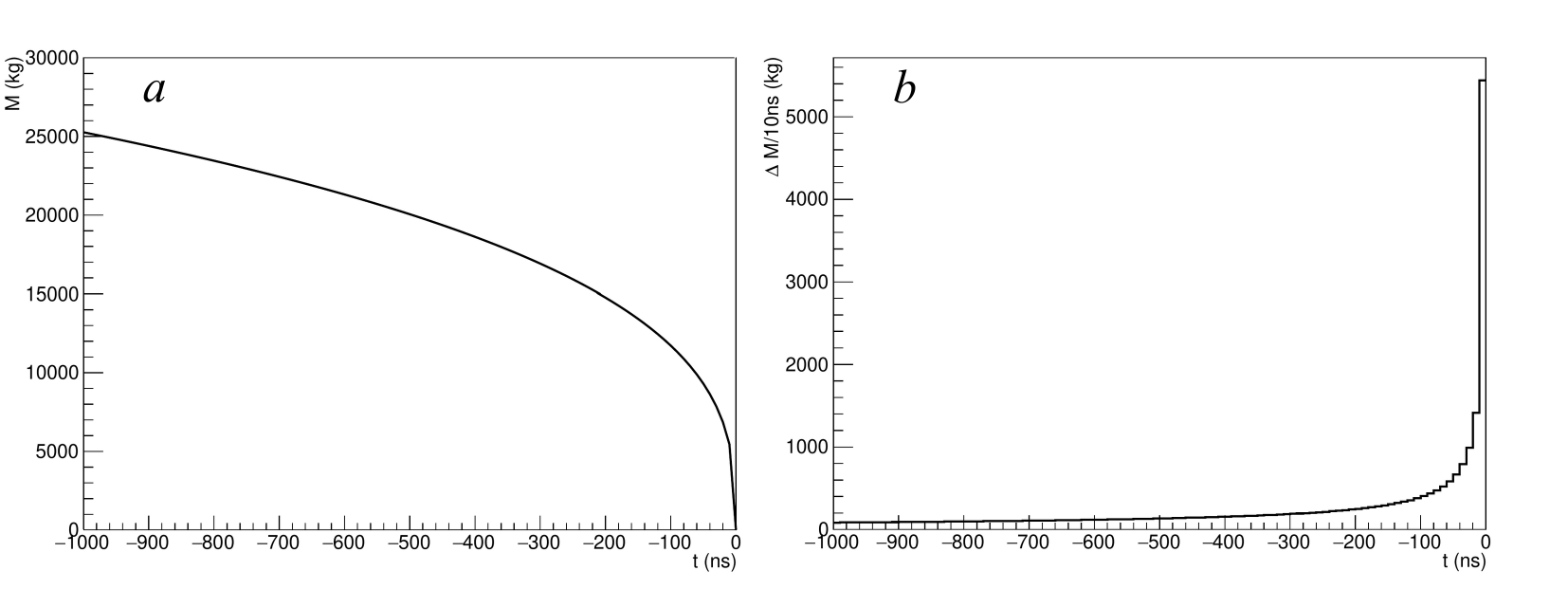}
 \end{center}
 \caption{$a$: The mass of a black hole as a function of the time to complete evaporation. $b$: Mass flow rate of the black hole for evaporation at 10\,ns intervals.}
 \label{fig:Tau-M}
\end{figure}

The \fig{Tau-M} shows how the mass of the black hole remnant changes over the last 1000\,ns before the final evaporation (panel $a$) and how the black hole mass is consumed for evaporation in time intervals of 10ns (panel $b$). The time interval is chosen to be equal to the integration time of the HiSCORE optical stations. The calculation is done with the formula \eq{MTau} under the assumption that $f_L = 100 = \const$. It can be seen that the evaporation ends with a very powerful explosion: during the last 10\,ns about 5000\,kg of mass is converted into the energy of ultrarelativistic particles. This is about 120 times the total luminosity of the Sun. This naturally leads to the idea of registering PBH evaporation events by detecting the products of the final explosion.


Many experimental studies have been carried out to search for evaporating black holes. Most of them search for gamma quanta of TeV energies that may be emitted during the final phase of the evaporation \cite{CYGNUS-1993,Tibet-1995,Whipple-2006,PETKOV-2008,PETKOV-2009A,PETKOV-2009B,Milagro-2015,VERITAS-2017,FERMI-2018,HAWC-2020,HESS-2023}. The criterion for the PBH evaporation event in these works is in most cases the grouping of detected gamma quanta in a single time interval, the duration of which varied in different experiments from 0.001\,s \cite{Milagro-2015} to 120\,s \cite{HESS-2023}, and the arrival of the events from a single point in the sky within the measurement errors, if the direction of the gamma ray arrival was determined at all. To date, this group of methods has obtained the best constraint on the frequency of PBH evaporation events as less than $2\times10^3\,\rm{pc}^{-3}\rm{yr}^{-1}$ \cite{HESS-2023}, and the LHAASO experiment plans to reach the $700\,\rm{pc}^{-3}\rm{yr}^{-1}$ \cite{LHAASO-2024} limit. The disadvantage of these methods is that they are purely statistical: accidental coincidence of background gamma rays of different origin by time and direction of arrival is possible. There is no unique signature of the PBH evaporation event, so the method only allows us to place upper bounds on the frequency of the events we are looking for.


Multimessenger-type methods seem more informative. The articles \cite{REES-1977,BLANDFORD-1977} considered the possibility of detectin a radio burster together with hard gamma quanta. The paper \cite{AMON-2015} notes that the simultaneous detection of neutrinos, gamma quanta, and cosmic ray particles such as neutrons and protons coming from the same direction gives an almost unambiguous signature of a PBH explosion (sources no further than 1\,pc from the Sun are considered, so neutrons can reach us due to a very large Lorentz factor, and very energetic protons travel almost in a straight line). In the paper \cite{PETKOV-2019} it is noted that a very informative signature of the explosion of a black hole remnant can be given by the simultaneous detection of the optical flare and gamma rays in the energy range from tens of MeV up to multi-TeV scale.

Models of black hole evaporation should be able to predict the energy spectrum of the emitted particles. There are models without a photosphere of BH \cite{MACGIBBON-1990,MACGIBBON-1991} and with a photosphere (fireball) \cite{HECKLER-1997A,HECKLER-1997B,DAGHIGH-2002}. The photosphere of an evaporating black hole is formed under the assumption that the primary particles of the Hawking radiation interact strongly with each other at sufficiently high temperatures, which should occur in the final stages of black hole evaporation. Moreover, at particle energies above the QCD phase transition temperature ($\sim100$\,MeV), the photosphere is expected to transition to the quark-gluon plasma state \cite{CLINE-1997,PETKOV-2004} and strongly scatter the primary particles of the Hawking radiation. Therefore, the particles emitted by the explosion of the black hole remnant will be predominantly secondary. Among the secondary particles there will be optical photons in some quantity, but it is assumed that the main mechanism of emission of soft photons will be the emission of photons of synchrotron nature in the external interstellar magnetic field \cite{REES-1977,BLANDFORD-1977} through the expanding envelope of the photosphere. The maximum of the spectrum of such radiation is expected near the frequencies $10^{14}$--$10^{16}$\,Hz \cite{REES-1977,BLANDFORD-1977,PETKOV-2019}, i.e. near the optical range. The time profile of such optical emission will of course not exactly match the curve in \fig{Tau-M}, but a very sharp spike near the time of final evaporation is expected. There will probably be some delay, so the main burst of ultra-hard gamma rays may arrive somewhat earlier than the optical flash. Unfortunately, we do not have detailed estimates of the time profiles of the various radiation types, taking into account the photosphere of an exploding black hole. One should also take into account that the profile of black hole evaporation in the final stages of the process can be modified by strong effects of quantum gravity, which cannot be taken into account in modern calculations, or, for example, by the presence of additional microscopic dimensions of space \cite{KOL-2002,KAVIC-2008,PETKOV-2019}. In particular, it follows that a detailed measurement of the time profile of black hole evaporation can provide a great deal of information for fundamental physics, far beyond what can be achieved at any promising particle accelerator.


The dead time of the HiSCORE optical stations is not more than 0.4 milliseconds. Using the formulas \eq{MTau} and \eq{kT} we can find that during the last 10 milliseconds of the black hole's existence its temperature will not be lower than a few PeV. The gamma-ray detection threshold for the HiSCORE array is no higher than 100\,TeV, therefore the HiSCORE array can in principle detect about a dozen gamma quanta with energies higher than 100\,TeV (probably -- multiPeV gamma quanta) during the last 10 milliseconds of the black hole's life before its explosion. All of these gamma rays, along with the optical flash at the time of the final evaporation, must be coming from the same point in the sky, within the measurement errors. Taken together, the gamma rays and the optical flash would provide an almost unambiguous signature of PBH evaporation. In other words, the HiSCORE array can detect black hole evaporation not only by the optical flash, but it is possible to realize the multimessenger mode of operation within one instrument, which dramatically increases the reliability of event identification.

3. Another type of astrophysical objects that can be detected and studied with the HiSCORE array are isolated stellar-mass black holes, the result of the evolution of massive stars. The critical feature for their detection is the stochastic variability of the radiation of the plasma accreting onto the BH \cite{SHVARTSMAN-1971,BESKIN-2005}. Such flares can be quite intense and can be detected by large instruments. In particular, in the approximation of adiabatic heating of plasma clumps formed in the accretion flow as a result of the development of instabilities, the amplitude of flares in the X-ray and gamma-ray regions can reach several percent of the steady state luminosity of the halo around the BH with their characteristic duration of $\tau \sim r_g/c \sim 10^{-4}\div10^{-5}$ seconds \cite{BESKIN-2005}. The temporal resolution of the order of microseconds, combined with the high sensitivity of large telescopes such as the Athena spacecraft, will allow the detection of this type of X-ray bursts, whose shape carries information about the properties of space-time near the event horizon \cite{BESKIN-2022}. It is extremely important that during the reconnection of magnetic field lines in current sheets, the main mechanism of particle acceleration near the event horizon, their maximum Lorentz factor can reach $10^4$--$10^5$ \cite{BESKIN-2005}. In these cases, for a remote observer, the flux density of the radiation of electron beams accelerated in current sheets when they are oriented along the line of sight increases as the inverse square of the Lorentz factor \cite{RYBICKI-1979}, and it becomes detectable by low sensitivity instruments, in particular the HiSCORE array. The steady-state luminosity of BHs with a mass of 10 solar masses is in the range of $10^{30}$--$10^{32}$\,erg/s \cite{Bisnovatyi-Kogan-1974,IPSER-1982,BESKIN-2005} depending on the BH velocity and the parameters of the interstellar medium. With beam Lorentz factors of $10^4$ to $10^5$, due to the concentration of energy in a very narrow beam, the apparent luminosity of the source at a distance of 100~pc can be in the range of $-7$ to $-4$ magnitudes for flares with an amplitude of 5\%. Such sources are well accessible for detection with HiSCORE. Since the number of BHs in the Galaxy is on the order of $10^8$ (see, e.g., \cite{WIKTOROWICZ-2019}), there are about 100 objects in the 100~pc radius region, and about 10 BHs are always in the field of view of HiSCORE. We emphasize that we should expect the detection of numerous outbursts, although an estimate of their frequency requires a detailed modeling of the accretion process.


4. If the source of the optical pulse is coherent, there is no limit to the size of the source to obtain a signal duration in the range of tens of nanoseconds. For example, according to the article \cite{PANOV-2021}, for a laser with an aperture of 1000\,m, for light with a wavelength of 0.53\,nm (green), removed at $10^4$ light years from the solar system, the diffraction radius of the light spot will be about 50 million kilometers. For the HiSCORE threshold of 3000 quanta/m$^2$/10\,ns we get that the laser energy should be about 9\,MJ per 10 ns to detect the pulse. Of course, we are talking about the laser of a hypothetical alien civilization. A laser aperture of 1000\,m can be achieved by using laser phased arrays. Laser phased arrays are devices long known on Earth, and the power of the required laser also corresponds to existing laser facilities used in inertial confinement fusion. Thus, no fancy equipment is needed to transmit information over kiloparsecs using lasers. The TAIGA HiSCORE array is well suited to detect optical pulses from such lasers, so if there are transmissions in the optical channel towards Earth, they can be detected. The large FOV of the HiSCORE array makes it possible to search for very rare events.

\section{Methods}

\begin{figure}
 \begin{center}
 \includegraphics[width=0.49\textwidth]{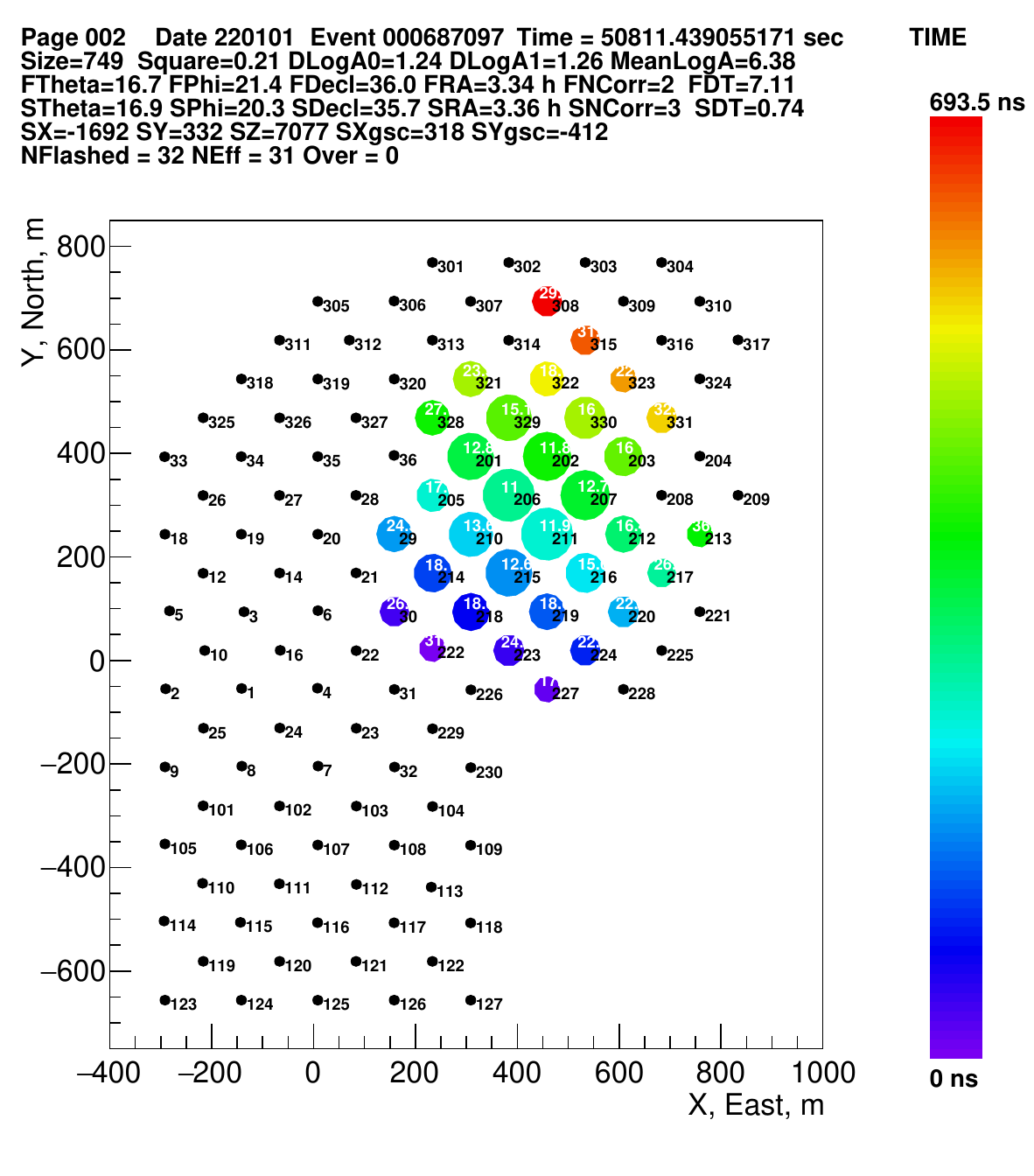}
 \includegraphics[width=0.49\textwidth]{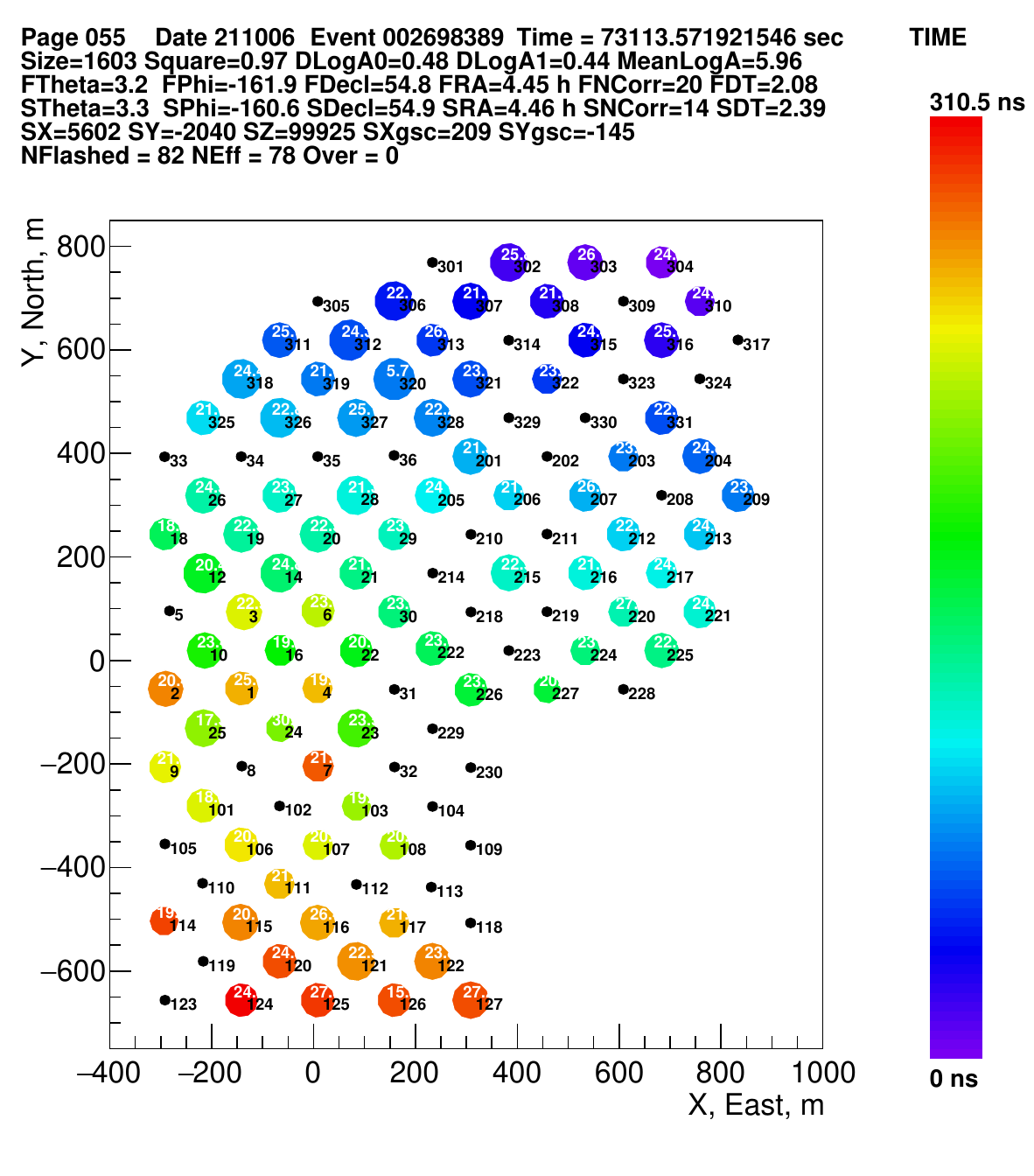}
 \end{center}
 \caption{Left panel: A typical EAS event. Right panel: One of the CALIPSO events. Both images are from the 2021--2022 data archive. In the figures: black small circles -- HiSCORE optical stations; black numbers -- internal numbering of HiSCORE stations; colored circles -- triggered stations; the color scale shows the relative time of station triggering, counting from the first triggered station (in nanoseconds); the size of the colored circle is proportional to the logarithm of the maximum amplitude of the signal of the triggered optical station; white numbers against the colored circles show the pulse duration of the optical station in nanoseconds. The upper part of each panel shows various characteristics of this event, some of which are directly obtained from the HiSCORE data archive (date, event number in the file corresponding to this date, time of event arrival (UTC, sec)). Other characteristics are the result of mathematical processing of the event.}
 \label{fig:Portraits}
\end{figure}

The initial basic idea of the method for extracting distant point optical transient events of astrophysical origin from the background of EAS events is explained in \fig{Portraits}. The left panel shows a typical EAS event. This is one of the events from the 2021--2022 season. The right panel shows one of the CALIPSO satellite events, also from the 2021--2022 season. It can be seen that EAS events typically have the shape of a compact ``spot'' with the amplitude of station signals increasing towards its center. The CALIPSO event should be similar to a signal from a distant point source, since the satellite passes over the HiSCORE array at an altitude of 700\,km -- much larger than the size of the array itself. It can be seen that there is more or less uniform illumination of the entire area of the array without a pronounced trend in the change of signal amplitudes. It is also clear that the optical station trigger times for distant transient events should be well approximated by the passage of a plane light front, while the response times for EAS events will be worse approximated by a plane front. These differences can be used to filter potential distant point optical transient candidates from the EAS background.


For the ``flatness'' criteria of the event in terms of optical station amplitudes and the quality criterion of the plane light front approximation to work well, it is necessary to perform a careful calibration of amplitudes and measurements of optical station trigger times. Such calibrations have already been performed during the processing of the 2018-2019 season \cite{PANOV-2021}, but they have been significantly improved in the present work.

\subsection{Calibrations}

\subsubsection{Amplitude calibration}


The purpose of amplitude calibration is to reliably compare the signal amplitudes of different optical stations in each event. The amplification factors of different optical stations are different, so in order to compare the amplitudes of different stations, corrections to the amplification factors must be made so that the amplification factors of all stations are effectively the same. It is important to note that adequate comparison of amplitudes from different stations does not require absolute calibration of optical station amplifications in the sense of precise coupling of amplitudes to absolute photon fluxes, as is required in the basic HiSCORE technique for EAS energy measurements.


The amplitude calibration technique described below repeats in its main features the technique already used for the Tunka-133 Cherenkov array \citep{TUNKA133-2020}, the predecessor of the HiSCORE array, but it has some peculiarities and has been implemented independently, specially for the problem of searching for optical transients.

\begin{figure}
 \begin{center}
 \includegraphics[width=\textwidth]{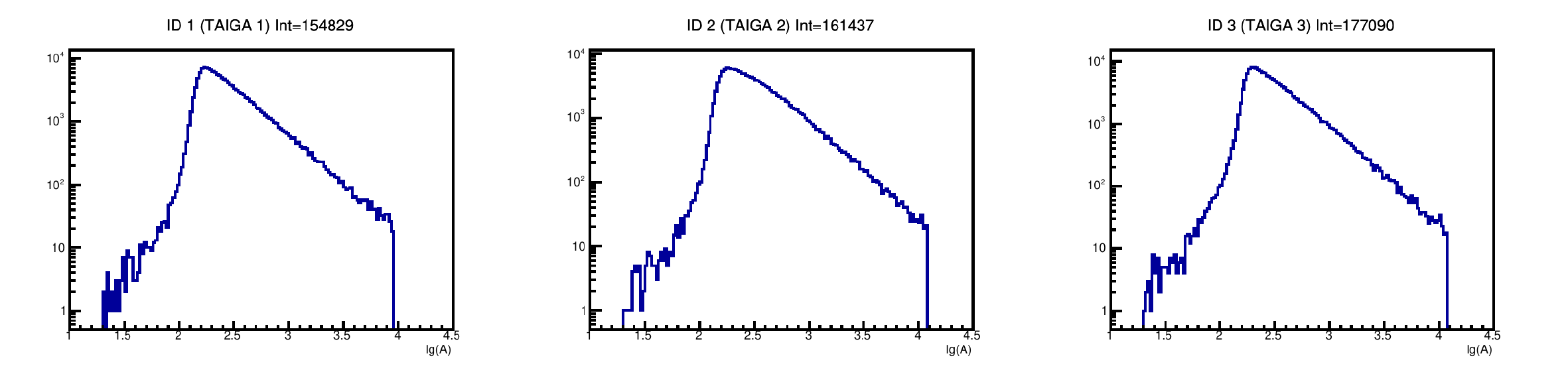}
 \end{center}
 \caption{Amplitude distributions at the optical stations with the numbers 1, 2, and 3, measured during a single night of HiSCORE observations.}
 \label{fig:Amplitudes}
\end{figure}

Each night of HiSCORE measurements in the data archive corresponds to a separate file. On different nights, the HiSCORE array settings may be slightly different, so amplitude calibration was performed separately for each night file. The idea behind amplitude calibration as follows. If all optical stations were operated in exactly the same way, then during a night of measurements all optical stations would produce exactly the same amplitude spectrum of detected signals, within the accuracy of statistical fluctuations, since all stations are in essentially the same conditions. The \fig{Amplitudes} shows the amplitude distributions obtained for three neighboring stations during one of the measurement nights. It can be seen that the amplitude distributions are slightly different, therefore different optical stations operate differently. The amplification correction should be done in such a way that the spectra of all stations are as similar as possible after the correction. This problem is solved as follows.


As can be seen from \fig{Amplitudes}, the spectra are close to a power law above the registration threshold. Each amplitude distribution is approximated by a power function of the form
\begin{equation}
 F(A) = B'\left(\frac{A}{A^*}\right)^{-\gamma'},
 \label{eq:FApower}
\end{equation}
where the constant $A^* = 1000$ is chosen for convenience of normalization. Technically, each distribution is fitted with functions of the form \eq{FApower} for all possible intervals in amplitude with interval lengths of at least one order of magnitude, and the best approximation is selected from all such approximations according to the criterion $\chi^2$. Then the mean values of all obtained parameters trough all working stations are calculated: $B = \langle B'\rangle$ and $\lambda = \langle \lambda'\rangle$. This results in a spectrum of the form \eq{FApower} averaged over all stations. It can be shown that in order to make the amplitude spectrum of each station equal to the average spectrum, the amplitudes must be transformed as:
\begin{equation}
 A = \left\{\left(\frac{A'}{A^*}\right)^{\displaystyle\frac{\gamma' - \gamma}{\gamma - 1}} \left[\frac{B(\gamma' - 1)}{B'(\gamma - 1)} \right]^{\displaystyle\frac{1}{\gamma - 1}} \right\} \times A'.
 \label{eq:FAcorrection}
\end{equation}
Applying the formula \eq{FAcorrection} to each measured signal amplitude is the implementation of the calibration. A direct check shows that after applying the  transformation \eq{FAcorrection} to the measured amplitudes, the amplitude spectra of all optical stations really become the same, except for the position of the thresholds (see Fig. ~2 in the article \cite{PANOV-2021}).


There is no way to equalize the amplitude thresholds as well, so after calibrating the optical stations, the thresholds will remain different. This means that even if the HiSCORE array is uniformly illuminated by a distant point source, it is possible that only some stations will be triggered while the sensitivity of some other stations remains below the threshold. However, as the analysis of the CALIPSO satellite events shows, even in such cases there is a uniform distribution of triggered stations over the array area, without a pronounced center near which the signal amplitudes would be larger than in the periphery.

\begin{figure}
 \begin{center}
 \includegraphics[width=\textwidth]{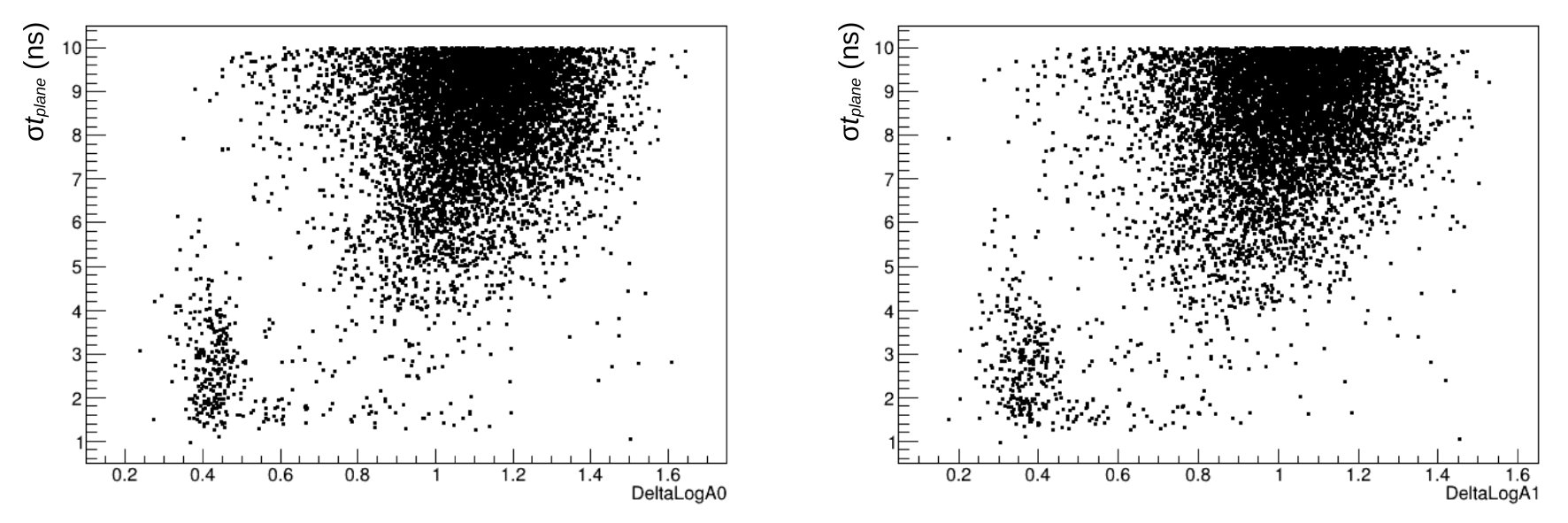}
 \end{center}
 \caption{Event distribution for DeltaLogA0 (left panel) and DeltaLogA1 (right panel) together with the RMS deviations of the station trigger times when approximated by the passage of a plane light front $\sigma t_{plane}$ (see section~3.2 for details). The data of the 2021--2022 season are shown.}
 \label{fig:DeltaLogA}
\end{figure}

To control the degree of ``flatness'' of the amplitude distribution in events, two parameters called DeltaLogA0 and DeltaLogA1 have been introduced. For each event, the DeltaLogA0 parameter represents the RMS deviation of the natural logarithm of the amplitude from the mean logarithm of the amplitude calculated from that event. To compute the DeltaLogA1 parameter, the entire distribution of natural logarithms of amplitudes in the plane of the HiSCORE array is approximated by the linear function of $X,Y$ (a plane), and DeltaLogA1 represents the RMS deviation of the natural logarithm of amplitudes from that plane. The \fig{DeltaLogA} for the 2021--2022 season shows part of the event distribution for DeltaLogA0 (left panel) and DeltaLogA1 (right panel). The RMS deviations of the station trigger times from the trigger times predicted by the optimal approximation of the light front by a plane light front (see section~3.2 for details) are plotted along the $Y$ axis. The groups of points in the lower left corner of the figures correspond to the events of the CALIPSO satellite, which should be close in structure to the events of distant point sources. It can be seen that the parameters DeltaLogA0 and DeltaLogA1 together with the RMS deviations of the station response times from the plane approximation $\sigma t_{plane}$ allow us to identify events of distant optical transients with a certain degree of confidence. However, as the analysis shows, some of the satellite events do not fall into these groups because of either too large deviations of $\sigma t_{plane}$ or too large deviations of the DeltaLogA0 and DeltaLogA1 parameters. Therefore, additional ways to filter out distant transient events are needed.

\subsubsection{Calibration of optical station trigger times}
\label{TimeCalibration}


Each optical station in each of the four clusters into which the stations are grouped in the HiSCORE array is connected to the cluster data center for data transmission via an optical cable. All four clusters are synchronized and data from the clusters can be downloaded to the HiSCORE array server. In order for the server to archive the correct values of the station trigger times, the signal delays in the measurement tracts must be known to an accuracy of at least one nanosecond. It is difficult to measure the signal delays with such accuracy in hardware, so residual uncertainties must be eliminated in a special online procedure to calibrate the optical station trigger times.


In our previous paper \cite{PANOV-2021}, such a calibration was performed using several hundred EAS events of ``maximum size'' (in terms of the number of simultaneously triggered stations) sampled over the entire 2018-2019 observation season. The statistics of this method were not very high, and there was no independent control of the quality of the calibration. Starting with the 2019-2020 season, there are substantially new possibilities for calibrating station trigger times.

\begin{table}
 \caption{All drone flights in HiSCORE observing seasons 2019--2020, 2020--2021, 2021--2022.}
 \label{tab:Drones}
 \begin{center}
  \begin{tabular}{|l|l||l|l|}
   \hline
   \multicolumn{2}{|c||}{Seazon 2019-2020} &    \multicolumn{2}{c|}{Seazon 2021-2022} \\
   \hline
    Date &        Time intervals (s, UTC) &      Date &        Time intervals (s, UTC) \\
   \hline
    2020-04-15  & 51200--53200 &                 2021-10-29 & 46500--48000;  54750--56750 \\
    2020-04-16  & 52100--53600 &                 2021-10-30 & 49520--50950                \\
   \cline{1-2}
   \multicolumn{2}{|c||}{Seazon 2020-2021} &     2021-11-01 &  45000--50000             \\
   \cline{1-2}
    Date        & Time intervals (s, UTC) &      2021-11-09 &  65500--66500               \\
   \cline{1-2}
    2021-02-02  & 55500--55800 &                 2022-01-28 &  47000--48400                 \\
    2021-04-14  & 61200--63300 &                 2022-02-22 &  60500--62100                 \\
    2021-04-15  & 68800--71500 &                 2022-03-04 &  57800--59300                 \\
   \cline{1-2}
     \multicolumn{2}{|c||}{} &                   2022-03-06 &  70400--72000                 \\
     \multicolumn{2}{|c||}{} &                   2022-03-07 &  67300--68700                 \\
     \multicolumn{2}{|c||}{} &                   2022-03-24 &  64900--66200                 \\
     \multicolumn{2}{|c||}{} &                   2022-04-28 &  64600--66000                 \\
     \multicolumn{2}{|c||}{} &                   2022-04-29 &  63200--64600;  56900--59400  \\
     \multicolumn{2}{|c||}{} &                   2022-04-30 &  56900--58300;  66900--68300  \\
   \hline
  \end{tabular}
 \end{center}
\end{table}


Since spring 2020, a drone with a LED on board flies over the HiSCORE array. The drone flies at an altitude of about 450 m and the LED emits light pulses with a frequency of about 10 Hz and duration of about 10 ns in the direction of the array. \tab{Drones} summarizes the drone flights during the three observation seasons 2019--2020, 2020--2021, 2021--2022.


The drone light pulse has a perfectly spherical front. This fact can be used to calibrate and correct the times of optical station triggers. For each drone pulse, the real times of the optical station triggers can be fit by passing through a spherical front at the speed of light. On the one hand, such a fit will show how the trigger times of the optical stations deviate from the optimally fitted spherical light front and thus how the trigger time of each station must be corrected to improve the quality of the fit. On the other hand, the position of the drone in the HiSCORE array coordinate system can be determined from the spherical light front fit. By reconstructing the drone coordinates pulse by pulse, it is possible to determine its trajectory, and by the quality of the trajectory reconstruction, it is possible to evaluate the quality of the reconstruction of the point-like source coordinates.


In fact, the time calibration procedure was implemented as follows. The main procedure of correcting the response times of the optical stations was not based on the approximation of spherical light fronts of drones, but on the approximation of light fronts of conventional EAS with the number of triggered optical stations not less than 12. EAS light fronts are not exactly spherical, although they are quite close to it, so the magnitude of the deviation of the actual trigger times from the EAS front fit in each case can also be used to determine the necessary corrections to the optical station trigger time. When processing a very large number of EAS fronts, the deviations of the front shape from exact sphericity are averaged, so that the averaged fit errors of each optical station provide well-defined corrections to be made to the hardware-defined station trigger times. The quality of the station trigger time correction is then independently verified by the quality of the drone spherical light front fits and the quality of the drone trajectory reconstructions.


At the first step of the described procedure, the quality of the optimally fitted light fronts will not be very good due to the errors in the times of the station triggers that have not yet been eliminated. So the first step estimation of errors of the optical stations trigger times will also be inaccurate. To improve the quality of the error estimation, the corrections to the station trigger times with error obtained at the first step should be introduced and the procedure should be repeated again. Now the front fit will be more accurate, and the new station error estimates will also be more accurate. This gives us an iterative procedure that must be repeated until it converges. In fact, it took 12 iterations in each of the three seasons 2019--2020, 2020--2021, 2021--2022 for the entire temporal calibration to converge.

\begin{figure}
 \begin{center}
  \includegraphics[width=\textwidth]{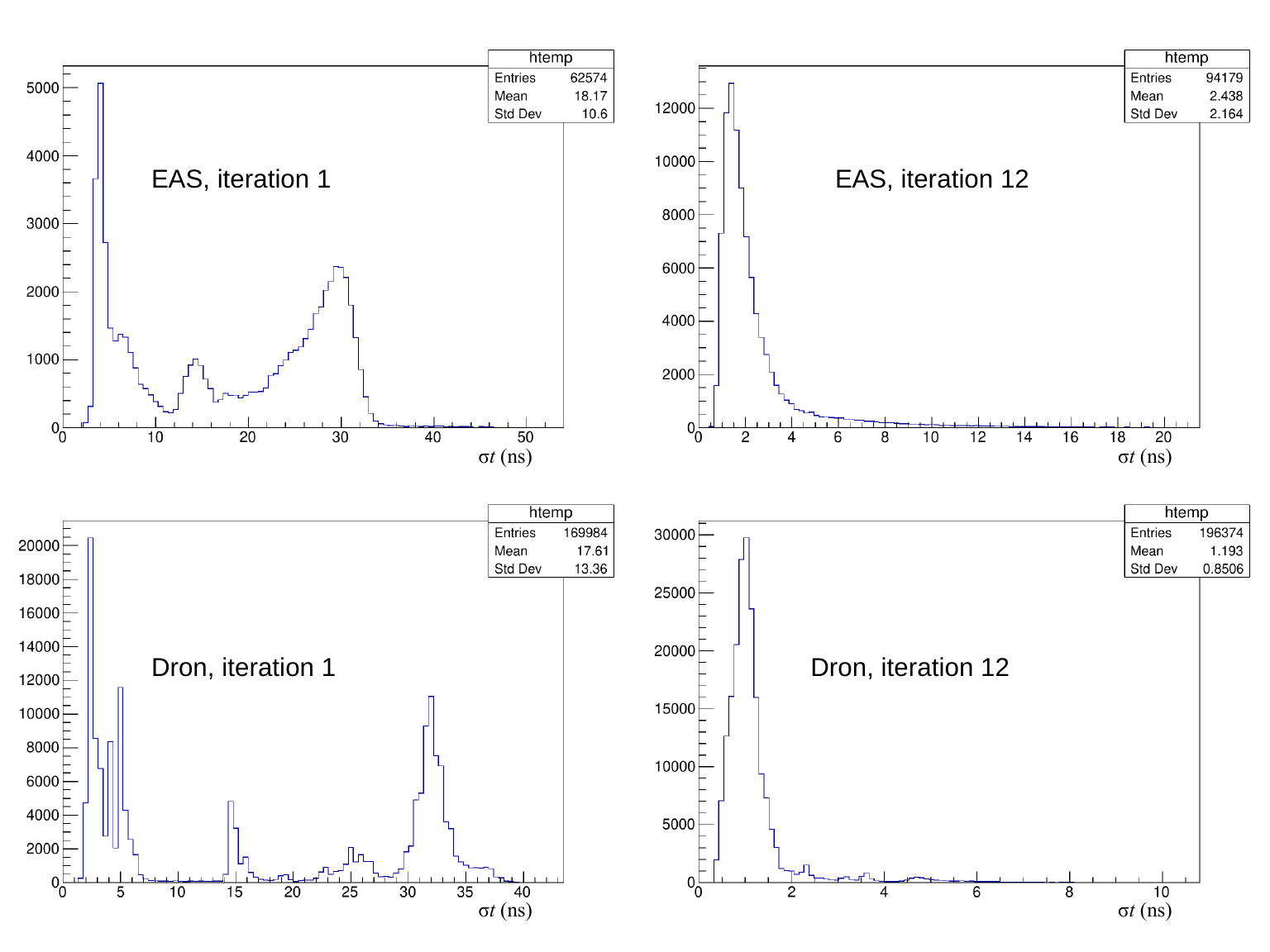}
 \end{center}
 \caption{Distribution of mean deviations of optical station trigger times from the optimal approximation by a spherical front. Top row -- approximation of EAS fronts. Bottom row --  approximation of drone LED fronts. Left column -- deviations after the first iteration, right column -- deviations after the 12th iteration.}
 \label{fig:TMeanError}
\end{figure}

The \fig{TMeanError} shows how the RMS deviation of station trigger times from the optimal approximation of the spherical light front has changed due to calibration, using the 2021--2022 season as an example. The top row of panels in \fig{TMeanError} corresponds to EAS events, the bottom row of panels in \fig{TMeanError} corresponds to drone events. The RMS deviation is the standard deviation calculated from the individual events. \fig{TMeanError} shows histograms of the distribution of this value. For both the EAS and drone events, we can see that the initial RMS errors in timing approach up to 40~ns, but after correction, the most likely error for the drone events is about 1~ns, and for the EAS events -- about 2~ns. The event error of the EAS event approximation is higher than the drone error because the EAS fronts are not perfectly spherical, so the approximation is not completely adequate. We can clearly see from the drone events that in addition to a narrow peak in the distribution of standard deviations with a maximum near 1~ns, the distribution has a weak tail that extends to about 6~ns. The existence of the tail is apparently explained by the fact that in addition to the systematic bias in the trigger times of the optical stations, which is compensated by the calibration performed, there are sometimes random jumps in the trigger times that are unpredictable and cannot be corrected completely. However, such jumps are relatively rare and therefore represent a solvable problem (see section~3.2).

\begin{figure}
 \begin{center}
  \includegraphics[width=\textwidth]{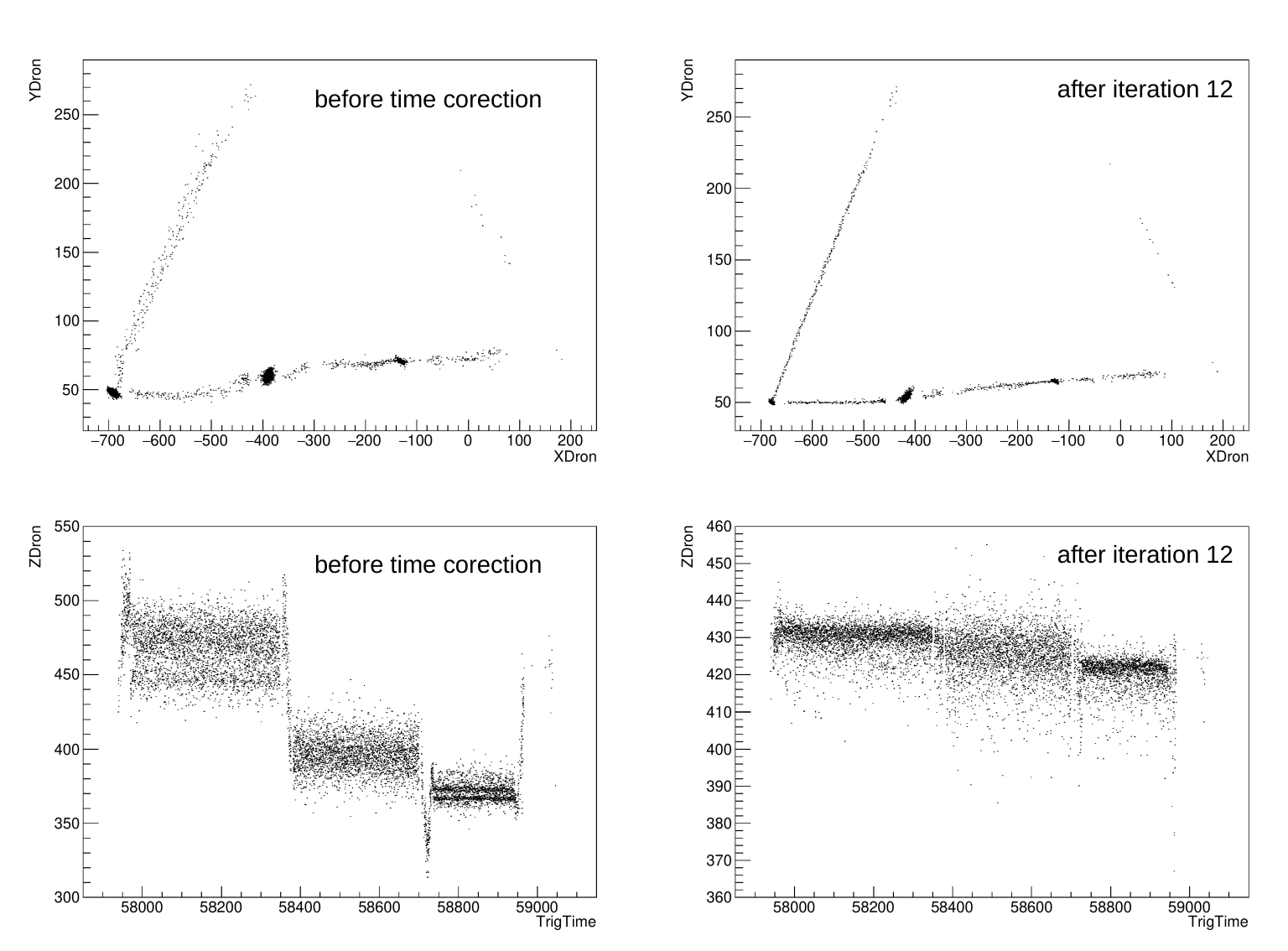}
 \end{center}
 \caption{Top row: projection of the drone flight trajectory on the XY plane. Bottom row: time dependence of Z coordinate. Left column: before corrections to the trigger times of optical stations. Right column: after corrections to the optical station trigger times (after the 12th iteration of the time correction procedure). All coordinates are measured in meters, TrigTime is measured in UTC seconds. The figure corresponds to the drone flight date 2022-03-04.}
 \label{fig:Dron}
\end{figure}

The \fig{Dron} shows how the accuracy of the drone trajectory reconstruction changes due to the calibration of the optical station trigger times. The left column of the panels shows the projection of the drone trajectory on the $XY$ plane and the dependence of the reconstructed drone coordinate $Z$ on the time before the calibration of the trigger times. The data refer to the drone flight date 2022-03-04. The right column of the panels shows the same trajectory after the trigger times calibration. It can be seen that the drone trajectory is already reconstructed before the trigger times calibration is performed, although the reconstruction errors are quite large. There are both large random errors (tens of meters) and systematic offsets of the same magnitude. After the calibration, all errors decrease dramatically. For example, we can see that the standard deviation in the determination of the drone height $Z$ is on a scale of 2--5~m, while the height itself is between 400 and 460~m.

\subsection{Approximation of HiSCORE events by plane and spherical light fronts}
\label{FrontAprr}


The problems of approximation of optical station trigger times by a plane or spherical light front are reduced to relatively simple nonlinear optimization problems. In the first case (plane front) the optimization is performed by three parameters, in the second (spherical front) -- by four parameters. The three optimization parameters for a plane front are the time when the front passes through the origin of the coordinate system and two angles that determine the direction from which the light front comes. The four optimization parameters for a spherical front are the time of emission and the source coordinates $X,Y,Z$. The time parameters of the optimization are not of interest and will not be discussed further.


The main problem in solving the approximation problems was not the technical details of implementing the optimization algorithms, but the need to optimize against the background of occasional large accidental errors (jumps) in the times of the station triggers. As mentioned in section~3.1.2, such jumps cannot be eliminated by pre-calibration of the optical station trigger times. The problem is further complicated by the fact that such accidental time jumps very often occur not in single optical stations, but in ``body groups'' of stations, which means that the station numbers with time errors are consecutive. Therefore, to correct such errors, not only large time errors in single stations, but also not very large errors but in ``body groups'' of the triggered stations should be filtered out.


The main idea of correcting for large accidental jumps in the times of station triggers when reconstructing the parameters of a plane or spherical wavefront is that it is not necessary to have very many triggered optical stations to reconstruct the front parameters. In principle, three stations are sufficient to reconstruct the parameters of a plane front, and four stations are sufficient to reconstruct the parameters of a spherical front. Therefore, in a case where there are quite a lot of triggered stations, it is better to get rid of a large number of stations that look unreliable and leave only the most reliable stations. If the remaining stations are two times as many as 3 stations for a plane front and 4 stations for a spherical front, the reconstruction of the background parameters will remain very reliable. Since the main part of the algorithms in this work considers only events where at least 12 stations are triggered, even a very strict filtering of unreliable stations does not lead to large losses in the number of reconstructedn events.


The same algorithm for filtering unreliable optical stations is used for plane and spherical fronts. In general, the algorithm consists of several iterations.


In the first iteration, the optimal light front approximation of all triggered optical stations is found. Then all stations are considered sequentially, and if it is found that in a group of stations with consecutive numbers, the total probability of deviation of trigger times in the same direction (i.e., either only in the positive direction or only in the negative direction) is less than three standard deviations, then the entire group of stations is removed and does not participate in subsequent iterations. Such stations we will call as \emph{bad staions}. There may also be only one station in the group, i.e., if the trigger time of one station deviates from the optimal fit by more than three standard deviations, then one station is removed. For the probability of a deviation in a group of stations to be less than three standard deviations, the deviations of each station do not have to be large: for example, it can be a deviation of only one RMS, but in several stations in a row and in the same direction. If no bad stations are found and removed in the first iteration, the process stops. This means that there were no large accidental time jumps (ejections), and the fit of the first iteration is the final result.


If a number of bad stations were discarded in the first iteration, the process is repeated. And so on. The procedure stops when no bad stations are found in any iteration. The fit result of the last iteration is the final result of the whole process. If the number of remaining stations on any iteration becomes less than 10, the process is terminated and the reconstruction of the light front parameters is declared to have failed. There are less than 0.1\% of EAS events where the attempt to find the optimal light front failed because the minimum number of ``good'' stations was violated. There are no such failures at all among the drone calibration events.

\begin{figure}
 \begin{center}
  \includegraphics[width=\textwidth]{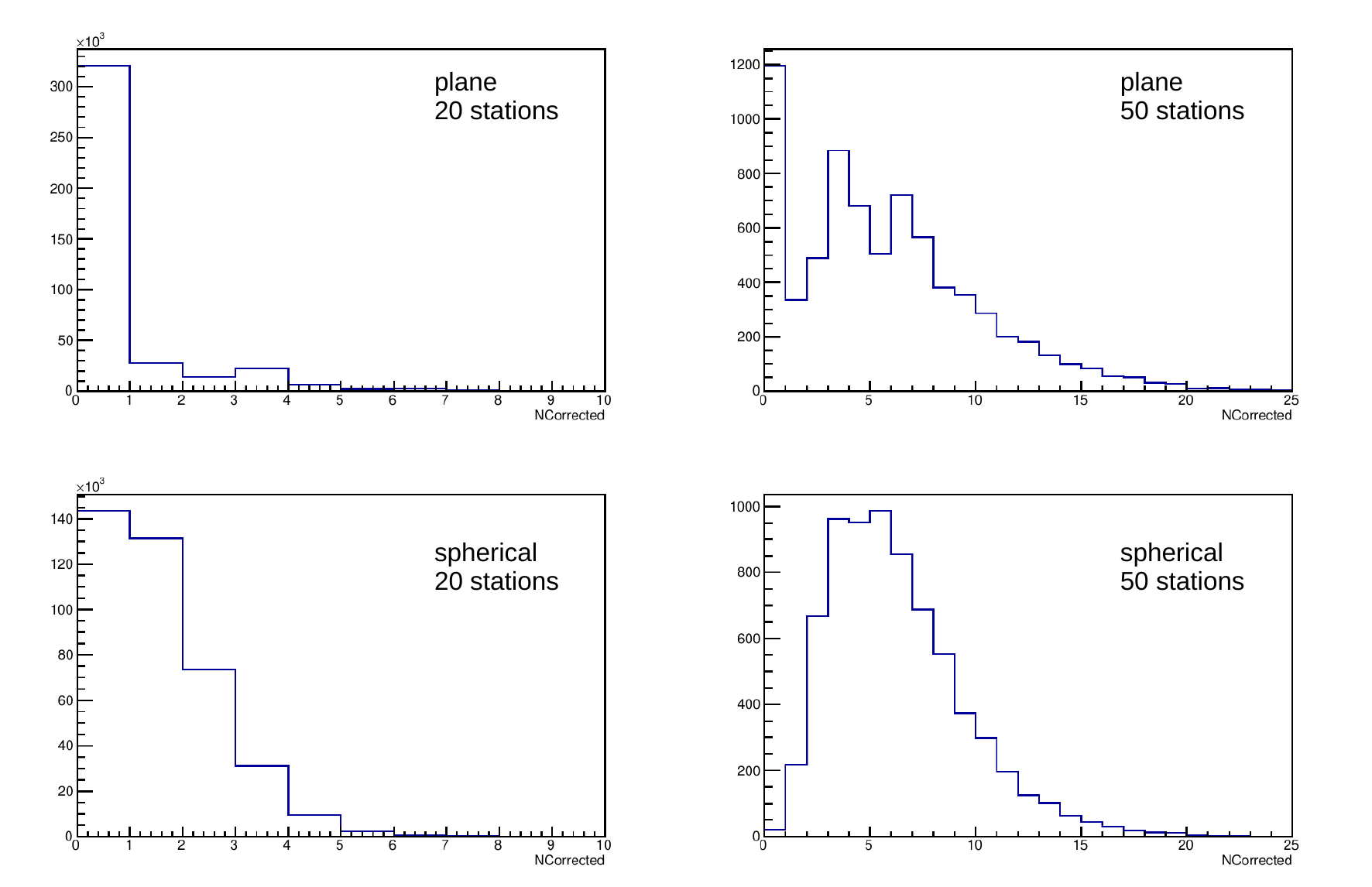}
 \end{center}
 \caption{The distribution of the number of ``bad'' optical stations (with possibly incorrect trigger times) removed from the event using the filtering algorithm described in the text (Section~\protect\ref{FrontAprr}). The data of 2021--2022 seazon are shown.}
 \label{fig:NCorrected}
\end{figure}

The histograms in \fig{NCorrected} show the distribution of the number of bad optical stations  removed using the filtering algorithm described above. The first row of histograms in \fig{NCorrected} corresponds to a plane front fit, the second row -- to a spherical front fit. For each type of front, the distributions are shown for events with an initial number of 20 triggered stations and 50 triggered stations. It can be seen that when the initial number of triggered stations is not very large (20), correction is often not necessary, i.e. the number of discarded bad stations is zero. When the number of triggered stations is large, there are only a few cases where there are no bad stations at all, and a correction is almost always necessary. We can also see from \fig{NCorrected} that the number of stations remaining after usage the filtering algorithm is usually large: for an initial 20 triggered stations, there are almost always at least 15 stations left at the last iteration, and for an initial 50 triggered stations, there are at least 30 stations left.


The direction to the source (the direction of the shower axis in the case of EAS) can be determined using either the plane front approximation or the spherical front approximation. To determine the direction to the source using the spherical front approximation, it is necessary to determine the position of the event axis in the plane of the HiSCORE array, since the algorithm gives the $X,Y,Z$-coordinates of the source (center of the spherical front) but not the direction to it. The direction to the source is the direction of the vector from the position of the event axis in the array plane to the center of the spherical front. What is the position of the event axis in the context of the algorithm based on the spherical front approximation?


In the original methodology, the approximation of the amplitude distribution of the optical stations using the spatial distribution function (SDF) obtained in the numerical Monte Carlo simulation of EAS was used to determine the position of the shower axis at the Tunka-133 and HiSCORE arrays \cite{TUNKA133-2020}. This approach focuses specifically on the reconstruction of EAS events, but in the present work the range of problems to be solved is broader. In fact, the curved light fronts we deal with in the present work are not only EAS light fronts, but also precisely spherical drone light fronts and curved CALIPSO satellite light fronts, which have a complex and not fully understood nature (see section~3.3.2). Therefore, the event axis position reconstruction algorithm should be more universal, if should not only focused on the EAS SDF. On the other hand, high precision in the reconstruction of the event arrival direction is not required for the algorithm using the spherical front, since this task is of an auxiliary nature in the context of this paper. Precision in direction reconstruction is required by an algorithm using a plane front reconstruction (see below). Therefore, a simpler and more universal method than in \cite{TUNKA133-2020} was used to reconstruct the position of the event axis together with the spherical front algorithm in the present work. The position of the event axis was determined by approximating the amplitude distribution of the triggered optical stations by a simple two-dimensional symmetric Gaussian function instead of EAS SDF.

\begin{figure}
 \begin{center}
  \includegraphics[width=\textwidth]{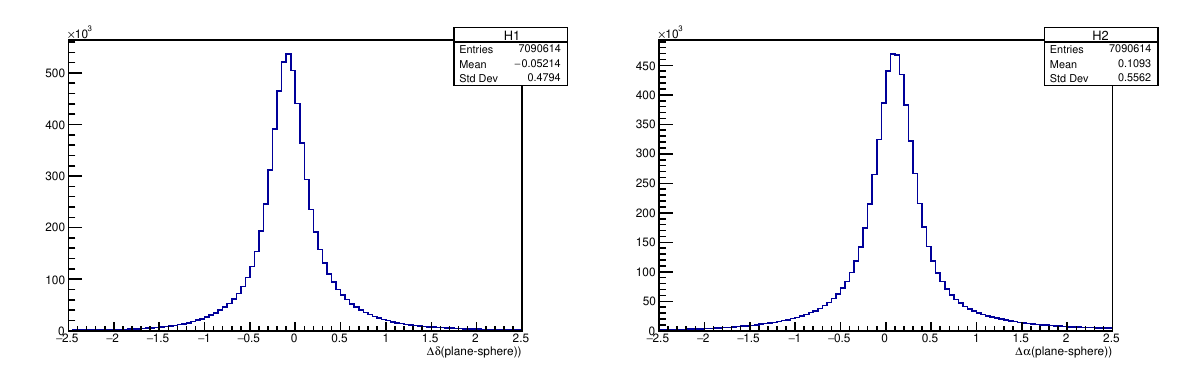}
 \end{center}
 \caption{Distributions of the difference in the determination of the direction to the source in the methods of approximating the light front by plane and spherical fronts. Histograms for declination $(\delta)$ and right ascension $(\alpha)$ are presented. For convenience, all angles are measured in degrees.}
 \label{fig:DeltaAngle}
\end{figure}

Due to the fact that the event direction recovery method for the spherical front approximation is quite approximate, we need to verify that the algorithm works reasonably well. The \fig{DeltaAngle} shows the distributions of the difference in the determination of the direction to the source in the methods of plane and spherical fronts for 2021--2022 data seazon. The angles are shown in the equatorial coordinate system: declination $(\delta)$ - right ascension $(\alpha)$. For convenience, all angles are measured in degrees. It can be seen that there is a systematic difference in the definition of the angles, although it is small: $0.05\degree$ for declination and $0.1\degree$ for right ascension. The RMS for the difference in angle determination is about $0.5\degree$. Therefore, in the general case, both algorithms -- spherical and plane-front -- work with an accuracy no worse than $0.5\degree$.

\begin{table}
 \caption{All CALIPSO transits over the HiSCORE array during the 2019--2020, 2020--2021, 2021--2022 data collection seasons.}
 \label{tab:CALIPSO}
 \begin{center}
  \begin{tabular}{|c|c|c|c|c|}
    \hline
    date  &  UTC time, sec & duration, sec & $\sigma\alpha$ (deg) & $\sigma\delta$ (deg)\\
    \hline
    \multicolumn{5}{|c|}{data seazon 2019--2020} \\
    \hline
    2019-10-07 & 70860 & 6.5  & $ 0.078\pm0.004$ & $0.082\pm0.004$\\
    2019-12-24 & 71110 & 7.5  & $ 0.062\pm0.003$ & $0.083\pm0.004$\\
    2020-01-19 & 71170 & 7.0  & $ 0.065\pm0.004$ & $0.112\pm0.006$\\
    2020-02-01 & 71190 & 7.7  & $ 0.060\pm0.003$ & $0.059\pm0.003$\\
    2020-02-27 & 71230 & 8.4  & $ 0.067\pm0.003$ & $0.064\pm0.003$\\
    \hline
    \multicolumn{5}{|c|}{data seazon 2020--2021} \\
    \hline
    2021-04-19 & 72440 & 10.4 & $ 0.043\pm0.002$ & $0.040\pm0.001$\\
    \hline
    \multicolumn{5}{|c|}{data seazon 2021--2022} \\
    \hline
    2021-10-06 & 73120 & 10.1 & $ 0.029\pm0.001$ & $0.057\pm0.002$\\
    2022-01-11 & 73590 & 9.1  & $ 0.032\pm0.001$ & $0.067\pm0.002$\\
    \hline
  \end{tabular}
 \end{center}
\end{table}

\begin{figure}
 \begin{center}
  \includegraphics[width=\textwidth]{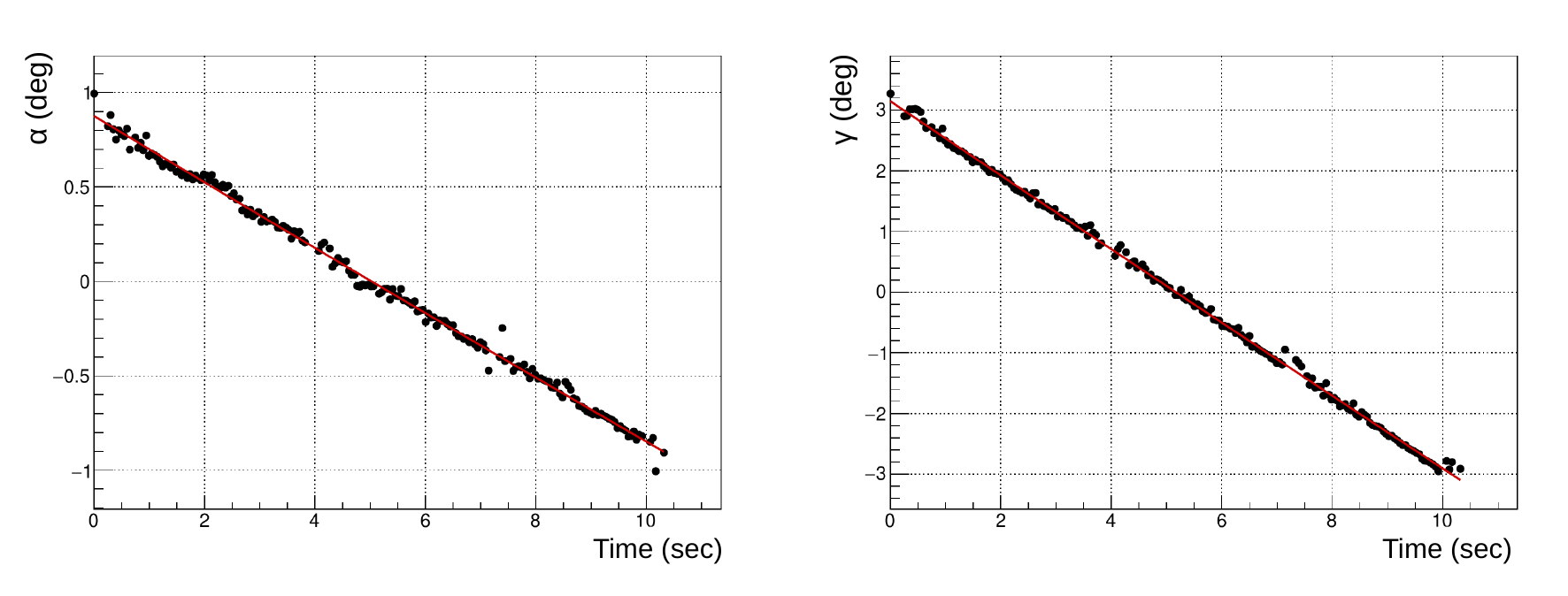}
 \end{center}
 \caption{Time dependence of the CALIPSO satellite's angular coordinates for the 2022-01-11 transit.}
 \label{fig:CALIPSO-Flight}
\end{figure}

CALIPSO transits over the HiSCORE array allow us to experimentally determine the accuracy of the reconstruction of the angular coordinates of events using a plane-front approximation of station trigger times. The \tab{CALIPSO} contains information on all CALIPSO transits in the seasons 2019--2020, 2020--2021, 2021--2022. The \fig{CALIPSO-Flight} for the 2022-01-11 transit shows the time dependence of the satellite's angular coordinates. The angular coordinates are chosen as angular coordinates whose origin is at the center of gravity of the trajectory, and the direction of one of the coordinate axes ($\delta$) coincides with the direction of the declination axis, and the other coordinate axis ($\alpha$) is perpendicular to the $\delta$-axis. Both coordinates are given in degrees. Since the transit time is only a few seconds and the satellite displacement during this time is only about two degrees, the time dependencies of the angular coordinates are very close to straight lines. Since the time of each CALIPSO event is measured very precisely, any deviation of the experimental points from straight lines in \fig{CALIPSO-Flight} is solely due to the accuracy of the angle measurements. By calculating the RMS of the deviations of the experimental points from the straight lines in \fig{CALIPSO-Flight} we find the RMS error of the coordinate measurement for each of the angles $\delta$ and $\alpha$ separately. The calculated errors are summarized in the last two columns of \tab{CALIPSO}. It can be seen that the errors are slightly different in different CALIPSO transits, but in general they are on a scale of $0.05\degree$.

\subsection{New signatures of the events of distant nanosecond optical transients.}
\label{NewSignatures}

\subsubsection{Event square filtration}


In our work \cite{PANOV-2021}, in which we analyzed the data of the first season of the search for nanosecond optical transients with the TAIGA-HiSCORE array (2018--2019), the main signatures for the selection of transient candidates were the ``event size'', the degree of flatness of the amplitude distribution in the event, and the quality of the approximation of the event by a plane light front. As mentioned in the introduction, this did not eliminate the EAS background in the region of zenith angles larger than $40\degree$. In the present work new, more efficient signatures were implemented.

\begin{figure}
 \begin{center}
  \includegraphics[width=\textwidth]{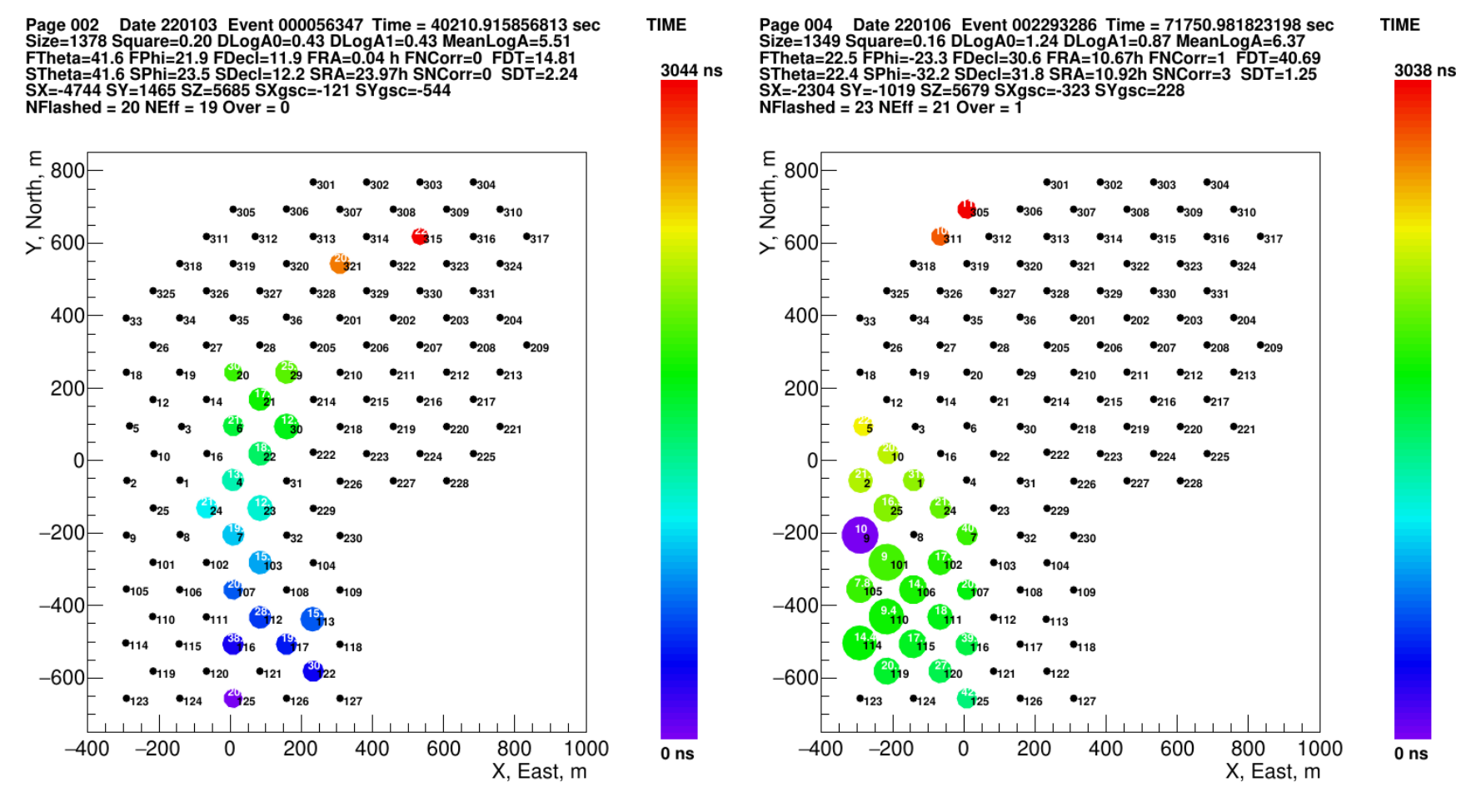}
 \end{center}
 \caption{Two examples of ``long but narrow'' events.}
 \label{fig:Long-Narrow}
\end{figure}

The event parameter ``event size'' refers to the maximum horizontal distance between the optical stations triggered in one event. When this parameter was introduced, it was thought that a large event size was a sign of uniform illumination of the entire array area in a single event. However, it turned out that there are often ``long but narrow'' events that have a large size but do not fill the HiSCORE array area uniformly. \fig{Long-Narrow} shows two examples of such events. Events of this type can result, for example, from the accidental overlapping of several weak EASs, or from the detection of the ``wing'' of a strong EAS whose axis is far outside the HiSCORE array.


In this paper, a parameter called ``event square'' (``EventSquate'' in plots and mathematics) is used instead of the ``event size'' signature. The ``event square'' refers to the area of the smallest convex polygon containing all the triggered stations of a single event. If we imagine a pin stuck into a horizontal surface instead of each optical station, then such a polygon is obtained by stretching an elastic rubber loop over all the pins. The maximum possible event area obtained with all 120 optical stations is taken as one. The areas of all other events are measured in fractions of the area of the maximum event, i.e. they can vary between 0 and 1. The ``long and narrow'' events discussed above have a small area, so they are effectively filtered out by a filter constructed using the ``event square'' parameter.

The probability of ``long and narrow'' events is small. For example, if we consider events of small area $\mathrm{EventSquate} < 0.2$ but large size $\mathrm{EventSize}>1200$\,m in the 2021--2022 data, the fraction of such events is $5\times10^{-5}$. However, in the search for very rare distant transient events, such ``long and narrow'' events provide a very significant background.

\begin{figure}
 \begin{center}
  \includegraphics[width=\textwidth]{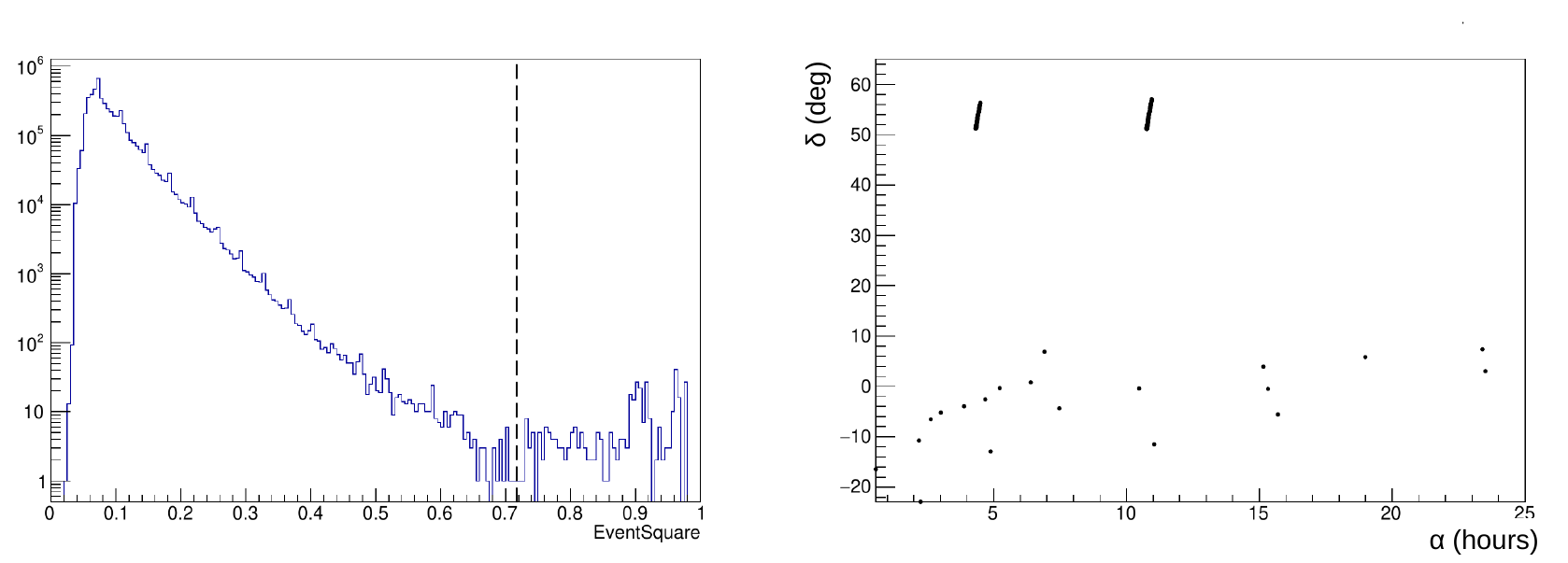}
 \end{center}
 \caption{Filtering with EventSquare parameter. Left panel: The EventSquare distribution for 2021-2022 seazon with the (rather weak) constraints $\sigma t_{plane} < 4$\,ns and $\mathrm{DeltaLogA0} < 1$. Right panel: All events above the threshold $\mathrm{EventSquare} = 0.72$ are plotted on the sky map in terms of equatorial coordinates.}
 \label{fig:EventSquare2}
\end{figure}

Filtering events using the EventSquare parameter proved to be very efficient. The left panel of \fig{EventSquare2}, for the 2021--2022 season, shows the EventSquare distribution with the (rather weak) constraints $\sigma t_{plane} < 4$\,ns and $\mathrm{DeltaLogA0} < 1$. There is a plateau of complex shape on the right side of the distribution. This plateau is almost exclusively represented by CALIPSO satellite events, and there is only a very small admixture of EAS events, all corresponding to large zenith angles and therefore small declinations. This can be clearly seen in the right panel of \fig{EventSquare2}. Here all events above the threshold $\mathrm{EventSquare} = 0.72$ are plotted on the sky map in terms of equatorial coordinates. Two short lines are visible at the top of the image -- these are CALIPSO events. Several dots scattered in the lower part of the map (small declinations) are EAS events. It can be seen that after selection by the condition $\mathrm{EventSquare} > 0.72$ there are only about two dozen unfiltered background EAS events from the entire 2021--2022 data collection season. However, additional filters are needed to filter them out as well.

\subsubsection{Altitude $Z$ filtration}
\label{ALTFILT}

\begin{figure}
 \begin{center}
  \includegraphics[width=\textwidth]{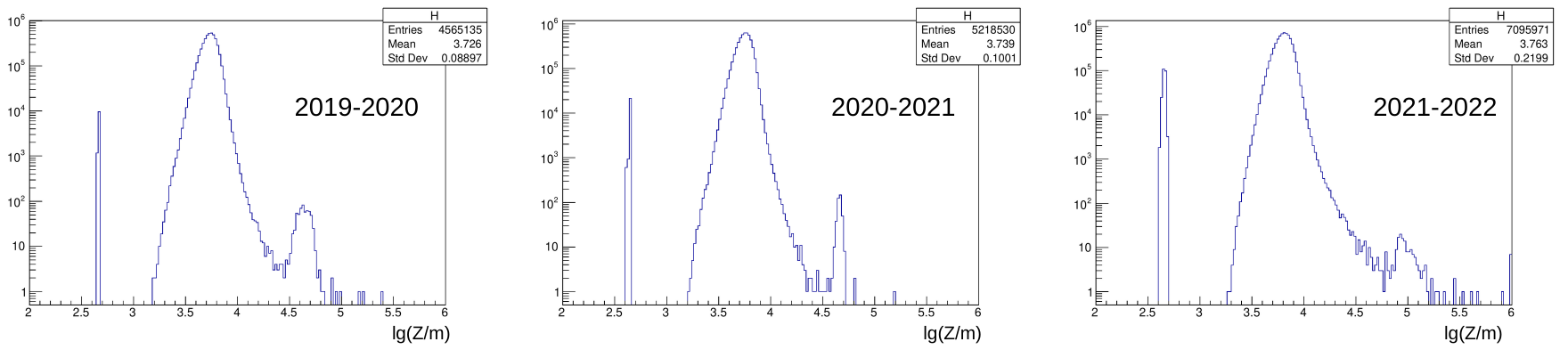}
 \end{center}
 \caption{Source altitude $Z$ (meters) distribution obtained by spherical light front approximation for the three seasons 2019--2020, 2020--2021, 2021--2022.}
 \label{fig:Z}
\end{figure}

Regardless of the assumed nature of the HiSCORE array event, the corresponding light front can be approximated by a spherical front and the effective coordinates of the light source can be found. \fig{Z} shows the distribution of the coordinate $Z$ of the events reconstructed by spherical front approximation for all three seasons 2019--2020, 2020--2021, 2021--2022. Three prominent details of the distribution can be seen in each plot. The narrow peak in the region between 400 and 500\,m corresponds to drone events. The high and broad peak in the middle of the distribution, between about 1500\,m and 30000\,m, corresponds to usual EAS events. The appearance of peaks on the right side of the distribution, at altitudes $Z$ greater than 30000\,m, is unexpected. Actaully, almost all events in the region of these peaks are CALIPSO satellite events. The CALIPSO satellite flies over the HiSCORE array at an altitude of about 700\,km, and one would expect the spherical front approximation to yield $Z$ estimates at this scale. The bending of the light front within the HiSCORE array area is sufficient to estimate the source altitude with an accuracy no worse than a hundred kilometers. However, the actual altitudes obtained are in the range of about 30\,km to 150\,km. Why so small? The reason is not well understood, but we can assume that it is a consequence of light scattering in the atmosphere. Because of scattering, a beam of light that initially has only diffraction divergence can begin to scatter in the upper atmosphere, and this can cause additional divergence. This additional divergence leads to a delay of a part of the light quanta with respect to the main front, which can be interpreted by the method of approximation by a spherical front as an effective light source at altitudes significantly lower than 700\,km. Another peculiarity of this effect is that this effective source is detected at different altitudes in different seasons. In the 2019--2020 and 2020--2021 seasons, the center of the distribution is located at an altitude of about 45\,km, while in the 2021--2022 season it is located at an altitude of about 90\,km and still has a complex shape. The reason for this behavior is still unclear. For example, it may be different atmospheric conditions in different seasons or some subtle peculiarities of the equipment operation.


In the 2021-2022 season data, there are about a dozen events for which the altitude of the source $Z$ was determined to be 1000\,km. This is not an error. The search range for the coordinate $Z$ in the optimization algorithm was the interval from 0 to 1000\,km, and when the right boundary of the interval was exceeded, the value $Z=1000$\,km was set as an indication that the optimal value of $Z$ is actually greater than 1000\,km. For true distant astrophysical transient events, the light arrives in a broad plane front rather than in relatively narrow beams as in the CALIPSO events. Therefore, the effects of light scattering in the atmosphere should not lead to an effective bending of the light front at the level of the HiSCORE array, and for such events we can expect exactly the determined height of 1000\,km.

\begin{figure}
 \begin{center}
  \includegraphics[width=\textwidth]{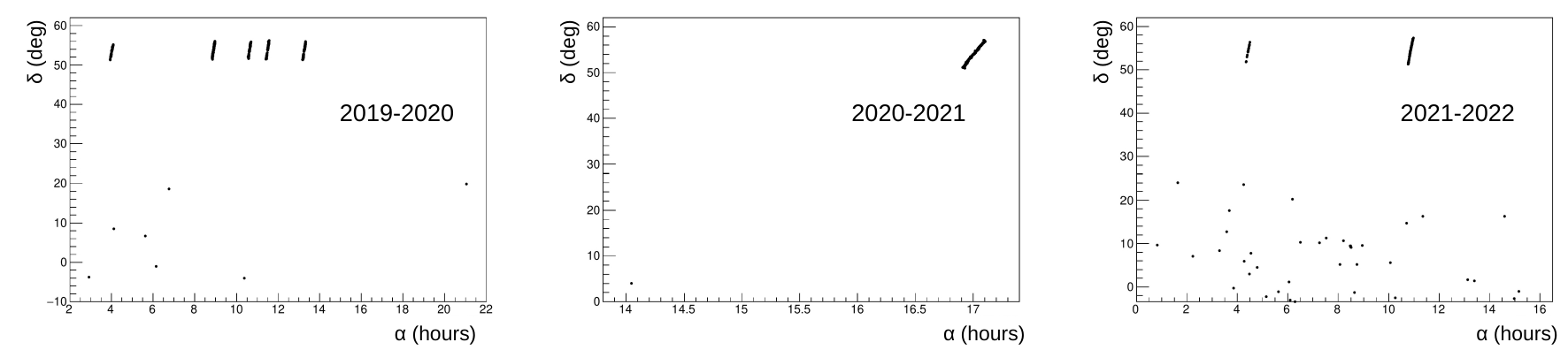}
 \end{center}
 \caption{Filtering of distant point-like sources by the altitude filter. The sky map for all events of the 2019--2020, 2020--2021, 2021--2022 seasons corresponding to the $Z$ distribution in \fig{Z} with the filters $\lg(Z/m)>4.45$, $\lg(Z/m)>4.50$, $\lg(Z/m)>4.81$ respectively.}
 \label{fig:ZSelection}
\end{figure}

Although the nature of the numerical value for the altitude of the CALIPSO source is not well understood, the correlation between the high altitude $Z$ and the nature of the source as a distant optical transient is obvious, so the source altitude $Z$ can be used as a signature to search for such events. \fig{ZSelection} in terms of equatorial coordinates shows on the sky map all events of the 2019--2020, 2020--2021, 2021--2022 seasons corresponding to the $Z$ distribution in \fig{Z} with the filters $\lg(Z/m)>4.45$, $\lg(Z/m)>4.50$, $\lg(Z/m)>4.81$ respectively.  It can be seen from \fig{ZSelection} that the CALIPSO satellite events (short lines at high declination) undergo such filtering, and only a fiew EAS events for each entire observation season undergo the filtering. Thus, filtering events by source altitude $Z$ is very effective in searching for distant optical transients.

\section{Results}
\label{RESULTS}


The basic idea used in this paper for the final filtering of distant optical nanosecond transient candidates is the combined use of the EventSquare filter and the event altitude $Z$ filter. The details of the implementation of this approach and the results obtained are presented below.


The filters for selecting distant point transients can be tuned by the following way. Using some data filtering tools unrelated to the EventSquare and $Z$ parameters, we select CALIPSO satellite events and treat these events as a sample of distant point light source signals. Then we use the selected CALIPSO events to see how EventSquare and $Z$ behave for them. Based on the analysis of the behavior of the EventSquare and $Z$ parameters for the CALIPSO events, it will then be possible to set filters based on these two parameters and then use these settings to search for candidates for distant optical transients of astrophysical origin in all HiSCORE data.


It is important that the filtering methods used for the initial selection of CALIPSO events are not used in the second (final) stage of the search for distant astrophysical transient candidates. That is, the filters for selecting satellite events and the filters for selecting transient candidates are completely independent of each other. It is not possible that the satellite event selection method could somehow influence the final selection of transient candidates.

\begin{figure}
 \begin{center}
  \includegraphics[width=\textwidth]{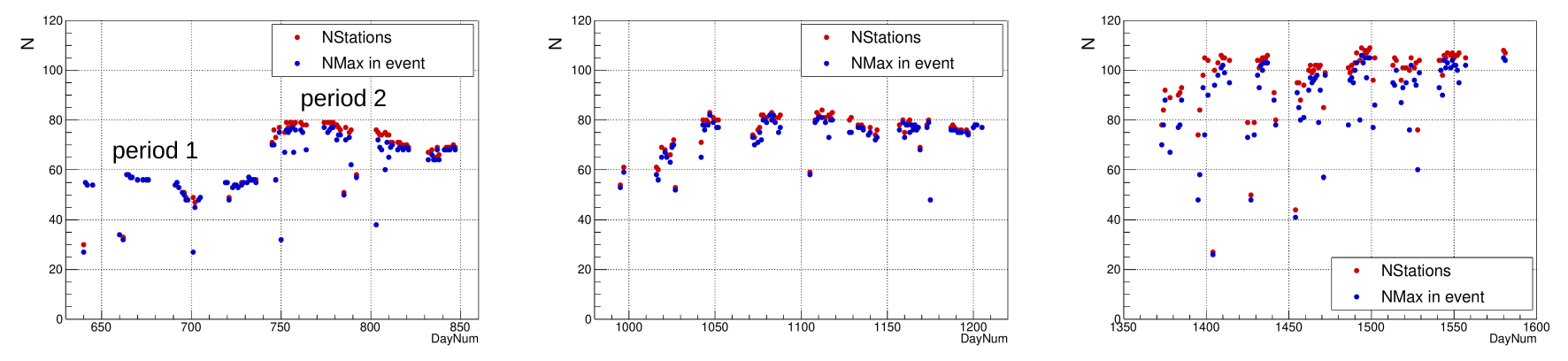}
 \end{center}
 \caption{Actual number of operating optical stations variations on individual nights of statistics collection. Time is measured in days counted consecutively from January 1, 2018. The parameter ``NStations'' shows the number of array stations triggered on a given night, and the parameter ``NMax in event'' shows the corresponding maximum number of stations triggered in a single event.}
 \label{fig:NStationsTime}
\end{figure}

The \fig{NStationsTime} shows for the HiSCORE seasons 2019--2020, 2020--2021, 2021--2022 how the actual number of operating optical stations varied with time on individual nights of statistics collection. Time is measured in days counted consecutively from January 1, 2018. The parameter ``NStations'' shows the number of array stations triggered on a given night, and the parameter ``NMax in event'' shows the corresponding maximum number of stations triggered in a single event in this night. It can be seen that the number of triggered stations increased on average from season to season, but there are numerous irregular variations in the number of triggered stations within this trend. A few particularly deep dips in the number of working stations are typically associated with either very short HiSCORE run times overnight (for various reasons: weather, etc.) or the shutdown of an entire cluster of stations for technical reasons. In the middle of the 2019-2020 season, we see a sharp increase in the number of working stations. This is due to the commissioning of about two dozen new stations into the array.


The filter settings for selecting distant optical transients depend on the number of working stations in the HiSCORE array. Since the number of working stations varies from night to night (see \fig{NStationsTime}), the filter settings should also vary. However, CALIPSO events are not available for calibration every night. Therefore, it was decided to calibrate over large observation periods characterized by a certain mean number of operating stations, each containing a CALIPSO transits over the HiSCORE array. The two halves of the 2019--2020 season, separated by the day number 740 (see \fig{NStationsTime}, these halves are labeled period~1 and period~2), the entire 2020--2021 season, and the entire 2021--2022 season were chosen as such periods. Separate filter settings were made for each of the four observation periods and optical transient candidates were searched for.

\begin{figure}
 \begin{center}
  \includegraphics[width=\textwidth]{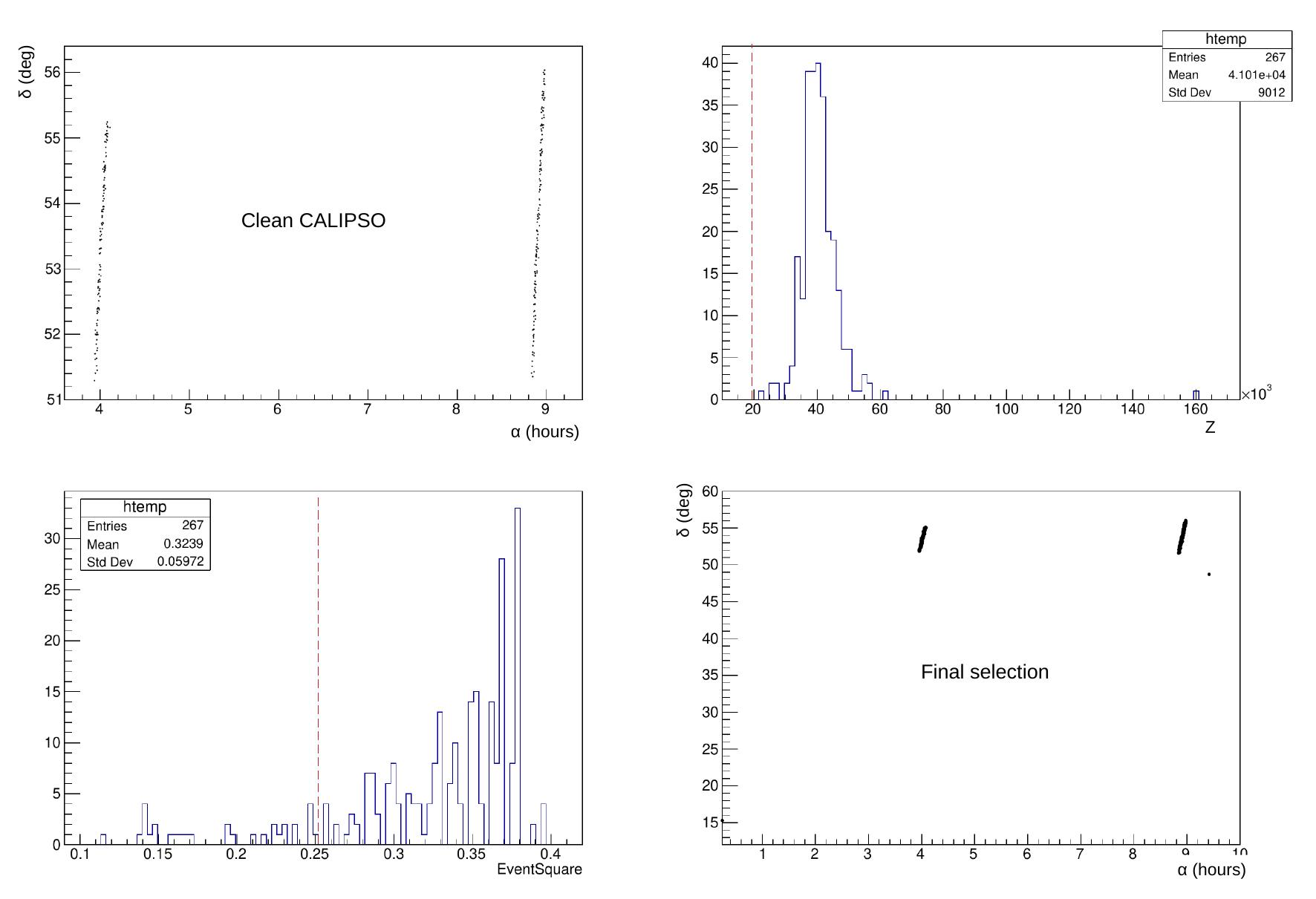}
 \end{center}
 \caption{The event selection procedure and its result for the 2019-2020 season, period~1.}
 \label{fig:Selection-2020-1}
\end{figure}

\fig{Selection-2020-1} shows the event selection procedure and its result for the 2019-2020 season, period~1. The upper left panel shows the CALIPSO satellite trajectory points during this period, filtered without the EventSquare and $Z$ parameters. The upper right panel shows the distribution of altitude $Z$ obtained from the CALIPSO events given in the upper left panel, and the lower left panel similarly shows the distribution of CALIPSO events by the EventSquare parameter. It can be seen that the distribution over EventSquare is quite broad and contains a tail on the left side with small EventSquare values. These points are caused by events close to the boundaries of the CALIPSO observing area, when the satellite events were almost outside the observability range. The maximum values of EventSquare are slightly below 0.4 -- which is significantly less than 1. This is due to the fact that in period~1 of the 2019--2020 array configuration, the number of optical stations in the HiSCORE array was significantly smaller than in the 2021--2022 season, for which the area normalization is performed. From the resulting distributions of $Z$ and EventSquare (\fig{Selection-2020-1}), we set thresholds for these parameters for the final selection of optical transient candidates: $Z_{thr} = 20$\,km and $\mathrm{EventSquare}_{thr} = 0.25$ (red dashed lines). These rather liberal selection thresholds allow all CALIPSO events by altitude $Z$ and cut off only the weakest CALIPSO events by EventSquare when they are at the visibility limit.

\begin{figure}
 \begin{center}
  \includegraphics[width=\textwidth]{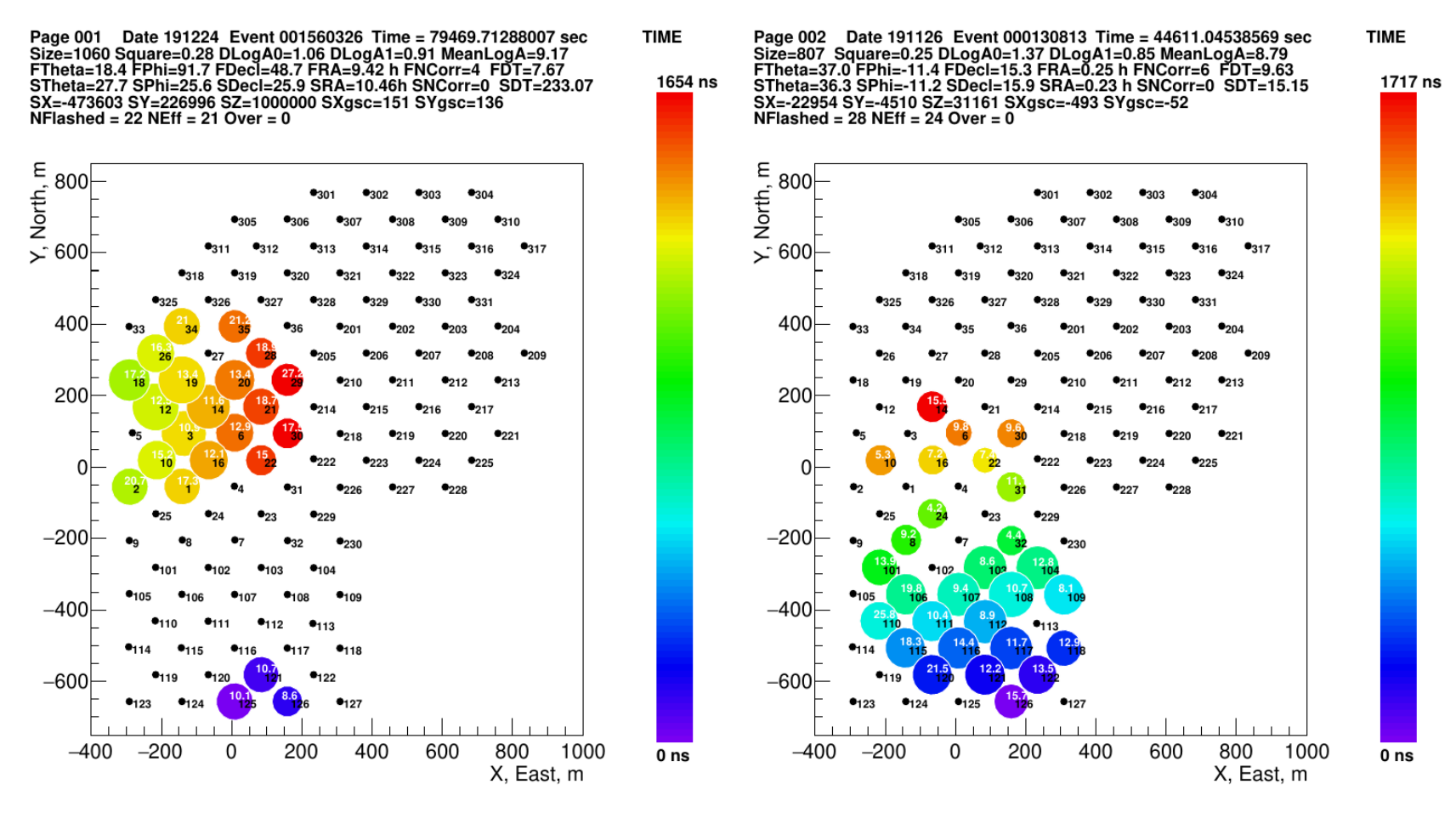}
 \end{center}
 \caption{Two events selected in 2019-2020 season, period~1 besides the CALIPSO events.}
 \label{fig:Selection-2020-1-Cand}
\end{figure}

The lower right panel of \fig{Selection-2020-1} shows the result of the selection of distant optical transient candidates for all period~1, 2019-2020. It can be seen that only two events were selected besides the CALIPSO events. These two events are shown in \fig{Selection-2020-1-Cand}. It can be seen immediately that these events are not at all like distant point tansients. In particular, they both have very poor approximation parameters by a plane light front. Note that in the final selection procedure described above, neither the quality of the plane front approximation nor any parameters other than EventSquare and $Z$ were used. If the approximation quality parameter had been used, the events would have been rejected and no events other than CALIPSO events would have been selected.

\begin{figure}
 \begin{center}
  \includegraphics[width=\textwidth]{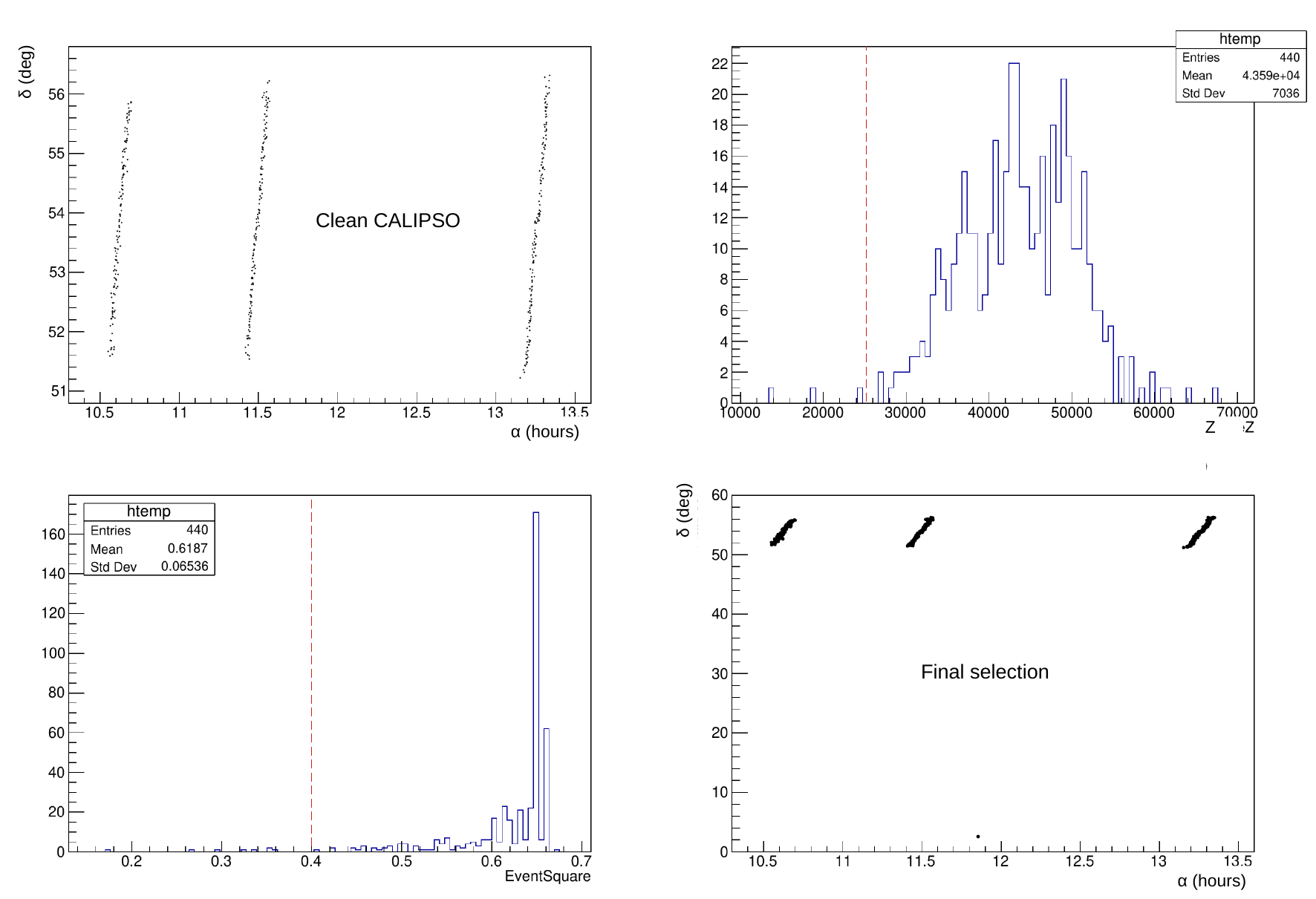}
 \end{center}
 \caption{The event selection procedure and its result for the 2019-2020 season, period~2.}
 \label{fig:Selection-2020-2}
\end{figure}
\begin{figure}
 \begin{center}
  \includegraphics[width=\textwidth]{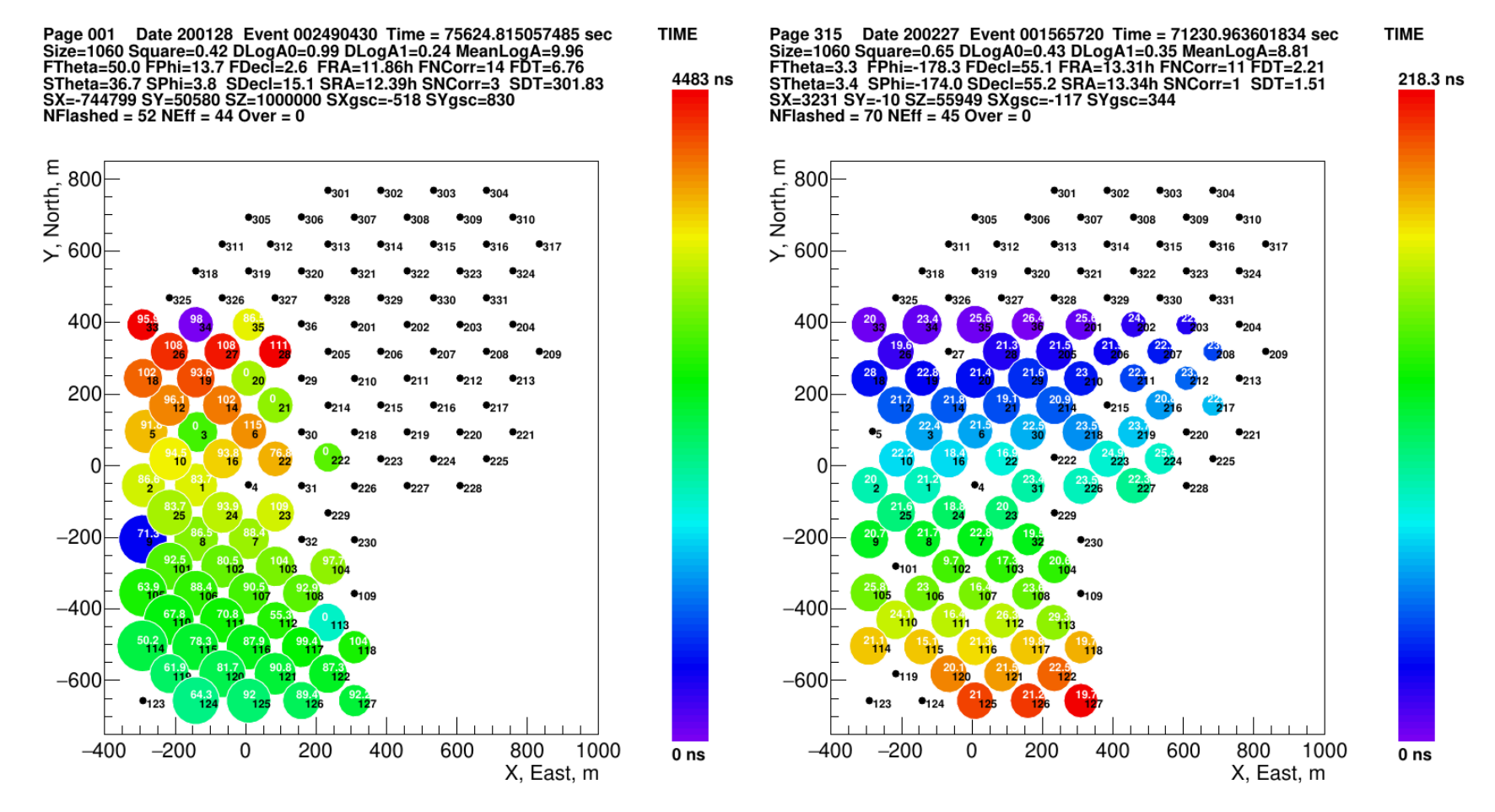}
 \end{center}
 \caption{Left panel: The only event selected in 2019-2020 season, period~2 besides the CALIPSO events. Right panel: One of the CALIPSO events, for comparison.}
 \label{fig:Selection-2020-2-Cand}
\end{figure}

The \fig{Selection-2020-2} shows the procedure and result of the selection of transient candidates for 2019-2020, period~2. The final selection used thresholds $Z_{thr} = 20$\,km and $\mathrm{EventSquare}_{thr} = 0.4$. Only one event that was not a CALIPSO event passed the filtering. This event is shown in the left panel of \fig{Selection-2020-2-Cand}. This event is indeed somewhat similar to the plane light front event, although there is still a tendency for the amplitude of the optical station signals to increase towards the lower left edge of the event. But this event is very different from the expected distant source event in that the region around $X=600$\,m, $Y=200$\,m is completely free of signal. For comparison, the right panel of \fig{Selection-2020-2-Cand} shows a typical CALIPSO event from the same time period and with approximately the same signal amplitude. The difference between the two events is striking. It has been verified that on 2020-01-28, when the selected event was detected, all three clusters of the HiSCORE array optical stations were working reliably, as well as on 2020-02-27, when the CALIPSO pass was detected. The fourth cluster of the HiSCORE array was not yet operational in period 2, 2019--2020 (the area at the top of the array). In other words, the absence of signals around $X=600$\,m, $Y=200$\,m of the selected event cannot be explained by the fact that one of the HiSCORE clusters was not working. Most likely the selected event represents the ``wing'' of a very strong EAS with a large zenith angle (about $50\degree$) and the axis located outside the array. The selected event cannot be considered to be a candidate for distant optical transients.

\begin{figure}
 \begin{center}
  \includegraphics[width=\textwidth]{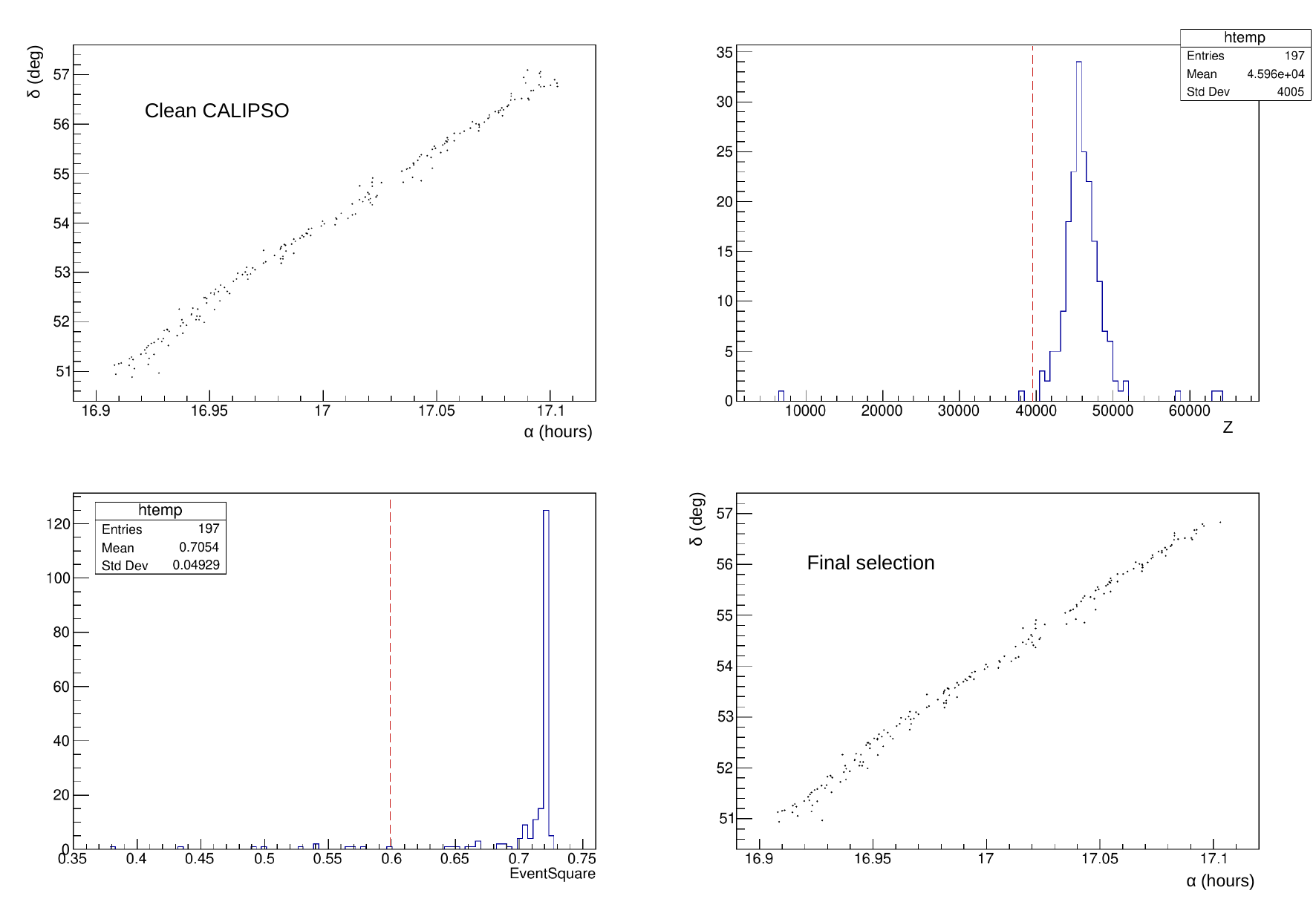}
 \end{center}
 \caption{The event selection procedure and its result for the 2020-2021 season.}
 \label{fig:Selection-2021}
\end{figure}
\begin{figure}
 \begin{center}
  \includegraphics[width=\textwidth]{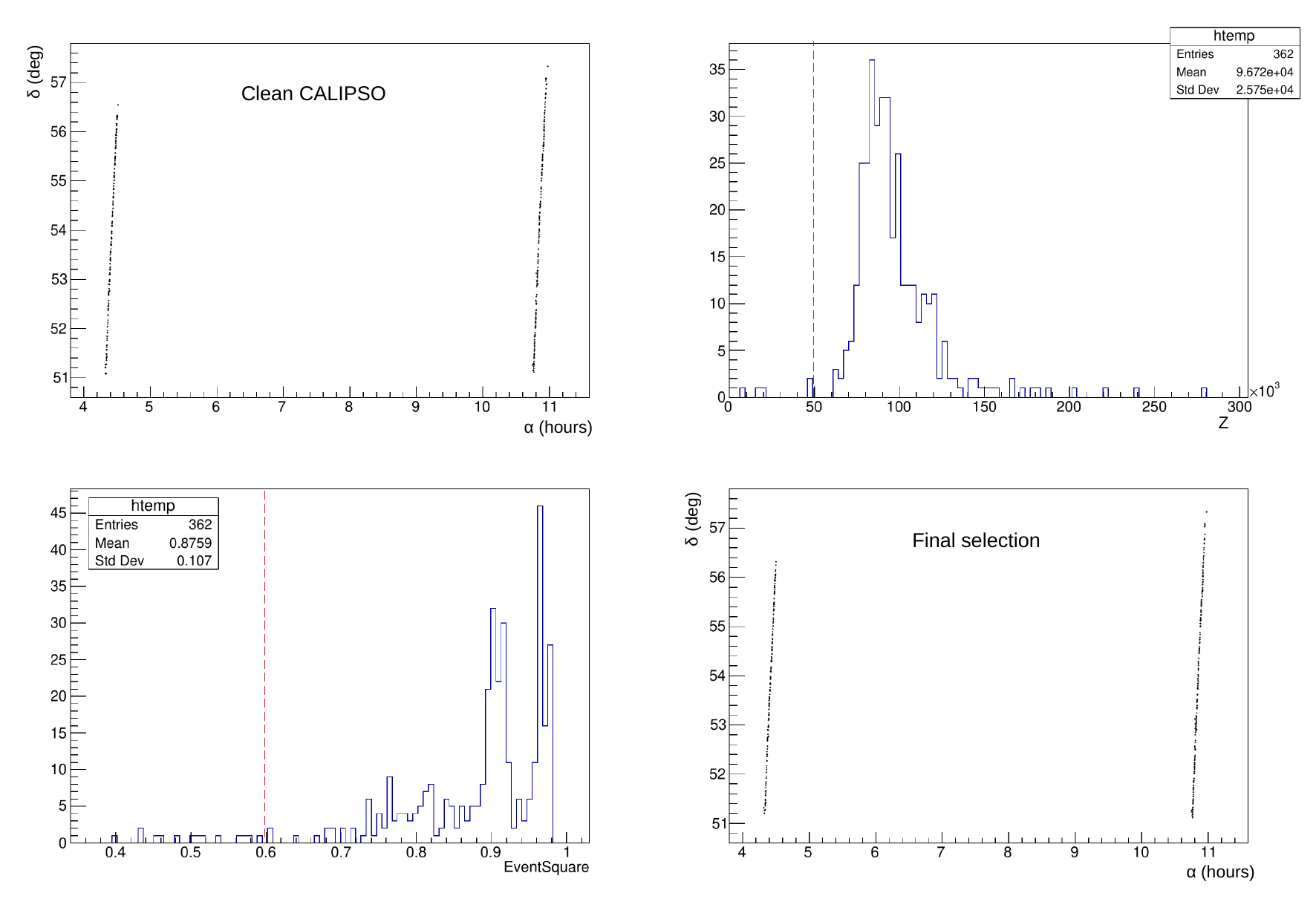}
 \end{center}
 \caption{The event selection procedure and its result for the 2021-2022 season.}
 \label{fig:Selection-2022}
\end{figure}

\fig{Selection-2021} and \fig{Selection-2022} show the procedures and results of transient candidate selection for the 2020--2021 and 2021--2022 seasons. The filter thresholds were set to $Z_{thr} = 40\,\mathrm{km}, \mathrm{EventSquare}_{thr} = 0.6$ and $Z_{thr} = 50\,\mathrm{km}, \mathrm{EventSquare}_{thr} = 0.6$, respectively. In both cases, the filters did not allow any candidates for distant optical transients other than CALIPSO events, in contrast to the 2019--2020 season. This result is easily explained. The HiSCORE array area and the number of operational optical stations in 2020--2021 and 2021--2022 increased compared to the 2019--2020 season (see \fig{NStationsTime}). As the area of the array grows, it becomes easier to distinguish spot-like EAS events from ``flat'' events of distant sources, therefore the background level will decrease as the array grows. This is exactly what is observed.

\section{Discussion}


Thus, no candidate for nanosecond optical transients of astrophysical origin has been detected in the last three seasons of observations. Nor was it detected in the first season of the HiSCORE search for nanosecond transients 2018--2019 \cite{PANOV-2021}. The HiSCORE exposure times, by data collection season, were: 2018--2019 -- 476~h; 2019--2020 -- 612~h; 2020--2021 -- 481~h; 2021--2022 -- 553~h. Using the FOV value of the HiSCORE array $\Omega \approx 0.6$\,ster \cite{PANOV-2021} we find the total HiSCORE exposure for all 4 observing seasons: $Exp = 1273\,\mathrm{ster}\cdot\mathrm{h}$. Since no transient candidates were detected, we get an estimate of the event flux density from above as $\sim1\times10^{-3}\,\mathrm{ster}^{-1}\mathrm{h}^{-1}$. The constraint applies to light flashes of 10\,ns duration or more with an energy flux density of at least $10^{-4}\,\mathrm{erg/s/cm}^2$. This upper bound on the event flux is the main result of this paper.

This work is supported by the Russian Science Foundation (grant 23-72-00019). The work of G.M. Beskin was carried out within the framework of the state assignment of SAO RAS, approved by the Ministry of Science and Higher Education of the Russian Federation.


\clearpage

\end{document}